\documentclass{nature}
\usepackage{hyperref}

\title{The quasi-linear nearby Universe}
\author{Yehuda Hoffman$^1$, Edoardo Carlesi$^{1,5}$, Daniel Pomar\`ede$^2$, R. Brent Tully$^{3}$, H\'el\`ene M. Courtois$^4$, Stefan Gottl\"ober$^5$, Noam I. Libeskind$^5$, Jenny G. Sorce$^{7,6,5}$, Gustavo Yepes$^8$,}

\usepackage{color}
\usepackage{graphics}
\usepackage[pdftex]{graphicx}
\usepackage{epstopdf}
\usepackage{epsfig}

\def  \LCDM{$\Lambda$CDM}

\def \kmsMpc {{\rm km s$^{-1}$Mpc$^{-1}$}}

\newcommand{\hmsun}{{\,\rm h^{-1}M}_\odot}
\newcommand{\hmpc}{{\,\rm h^{-1}Mpc}}
\newcommand{\hkpc}{{\,\rm h^{-1}kpc}}
\def\br{{\bf r}}

\def\bv{{\bf v}}

\begin{document}

\maketitle

\begin{affiliations}
 \item Racah Institute of Physics, Hebrew University, Jerusalem 91904, Israel
 \item Institut de Recherche sur les Lois Fondamentales de l'Univers, CEA Universit\'e Paris-Saclay, 91191 Gif-sur-Yvette, France
 \item Institute for Astronomy (IFA), University of Hawaii, 2680 Woodlawn Drive, HI 96822, USA
 \item University of Lyon; UCB Lyon 1/CNRS/IN2P3; IPN Lyon, France
  \item Leibniz Institut f\"ur Astrophysik, An der Sternwarte 16, 14482 Potsdam, Germany
  \item Universit\'e de Strasbourg, CNRS, Observatoire astronomique de Strasbourg, UMR 7550, F-67000 Strasbourg, France
  \item Univ Lyon, Univ Lyon1, Ens de Lyon, CNRS, Centre de Recherche Astrophysique de Lyon UMR5574, F-69230, Saint-Genis-Laval, France
 \item Departamento de F\'{\i}sica Te\'orica and CIAFF,  Universidad Aut\'onoma de Madrid, Cantoblanco 28049, Madrid Spain
 \end{affiliations}

\parindent 0pt

\begin{abstract} 

The local 'universe' provides a unique opportunity for testing cosmology and theories of structure formation. To   facilitate this opportunity we present  a new method for the reconstruction of the quasi-linear matter density and velocity fields from galaxy peculiar velocities and apply it to the Cosmicflows-2 data. The method consists of constructing an ensemble of cosmological simulations, constrained by the standard cosmological model and the observational data. The quasi-linear density field is the geometric mean and variance of the fully non-linear density fields of the simulations. The main nearby clusters (Virgo, Centaurus, Coma), superclusters (Shapley, Perseus-Pisces) and voids (Dipole Repeller) are robustly reconstructed. 

Galaxies are born `biased` with respect to the underlying dark matter distribution.  Using our quasi-linear framework  we demonstrate that the luminosity-weighted density field derived from the 2M++ redshift compilations is non-linearly biased with respect to the matter density field. The bias diminishes in the linear regime.

\end{abstract}

\bigskip



{\bf Introduction}\newline
Our local  neighborhood is a special arena for studies of large scale structure and galaxy formation. The high precision and the wealth of observations of the nearby structure can provide stringent tests of evolution models. However, the comparison of observations with theory is confronted with the issue of cosmic variance. The constraints provided by the local variance manifested in our neighborhood must be within the framework of an ensemble of possibilities described by a viable cosmological model. 

Attempts to uncover the local  density and velocity fields from observational data are not new. Studies can be classified according to the data employed  (redshift surveys or peculiar velocities), the dynamical range (linear or non-linear) and their statistical nature (mean fields or  individual constrained  realizations).   Early on, peculiar velocities surveys were used to uncover the local structure, starting with POTENT\cite{1990ApJ...364..349D} and then followed by the Wiener filter\cite{1995ApJ...449..446Z} (WF) and constrained realizations (CRs)\cite{1991ApJ...380L...5H}   method and other 
 Bayesian reconstruction methods\cite{1993ApJ...415L...5G,1999ApJ...520..413Z,2014Natur.513...71T,2015MNRAS.452.1052L,2016MNRAS.457..172L,2017NatAs...1E..36H}.  These velocity based reconstruction schemes are formulated and applied in the linear regime. About the same time the IRAS redshift survey was employed to reconstruct the local density and velocity fields\cite{1988lsmu.book..255S,1991ApJ...379....6N,1994ApJ...423L..93L,1993ApJ...412....1D,1995MNRAS.272..885F,1995PhR...261..271S}. 
A variety of studies probe the local non-linear large scale structure by means of constrained simulations  based on either  redshift surveys\cite{1996ApJ...458..419K,1998ApJ...492..439B,2002MNRAS.333..739M,2014ApJ...794...94W} or peculiar velocities data\cite{2000ASPC..201..169V,2016MNRAS.460.2015S,2016MNRAS.455.2078S,2010arXiv1005.2687G,2009LNP...665..565H,2008MNRAS.386..390H,2003ApJ...596...19K,2002ApJ...571..563K}.

The sampling of the non-linear local structure by means of constrained simulations has been recently extended to a Bayesian estimator of the  quasi-linear (QL) density, evaluated by sampling over an ensemble of redshift survey constrained realizations\cite{2012MNRAS.tmpL.528K,2016MNRAS.455.3169L,2018MNRAS.474.3152D}. It is the aim of our paper to present a new method for the estimation of the  QL matter density field from surveys of galaxy peculiar velocities and to apply the method to the grouped Cosmicflows-2 (CF2) data  of peculiar velocities\cite{2013AJ....146...86T}.   

Reconstructions of our local patch of the Universe from alternatively redshift surveys or peculiar velocity compendia have pros and cons.  These two data types provide different windows into the local large scale structure. 
Redshift surveys sample the distribution of galaxies that meet certain criteria, while peculiar velocities directly but sparsely probe the underlying matter density field.
The comparison of the two sheds light on  the still quite poorly understood processes of galaxy formation.

Here, for the first time, the QL density field is recovered from peculiar velocities. 
The aims of the paper are the introduction and presentation of: 1.   a new methodology for the QL estimation from peculiar velocities; 2.   an overview of the resulting local density field;
3. a comparison of the estimated local QL matter density field with the observed local galaxy distribution and thereby  probe  the  galaxy bias in the local `universe'. 

{\bf Theoretical Background}

One of the main tenets  of the standard model of cosmology is that structure has emerged via gravitational instability from a primordial Gaussian perturbation field, whose power spectrum is encoded in the temperature anisotropies of the Cosmic Microwave Background (CMB) radiation\cite{1980lssu.book.....P,2008cosm.book.....W}. 
The growth of structure  is accompanied  by motions -   hence the density and flow fields are coupled by the equation of continuity. 
It follows that  observations of  peculiar velocities of nearby galaxies can be used to recover the local underlying matter density field.  
Indeed the   WF/CRs tools have been applied to the   grouped CF2 data  and the linear local density and velocity fields were  reconstructed\cite{2014Natur.513...71T,2015ApJ...812...17P,2017NatAs...1E..36H}.  The linear WF/CRs algorithm is extended here by means of fully non-linear constrained simulations\cite{2009LNP...665..565H,2014NewAR..58....1Y,2014MNRAS.437.3586S,2016MNRAS.455.2078S,2016MNRAS.460.2015S}, namely numerical cosmological simulations starting from initial conditions constrained by the grouped version of the CF2 data of peculiar velocities. The road taken   in the linear regime from constrained realizations of Gaussian fields\cite{1991ApJ...380L...5H}   to the Bayesian conditional mean field, namely the Wiener filter estimator\cite{1995ApJ...449..446Z}, is paralleled here by  going from individual constrained simulations\cite{2013MNRAS.430..888D,2013MNRAS.430..912D,2013MNRAS.430..902D,2016MNRAS.455.2078S,2014MNRAS.437.3586S} to the QL estimator which approximates the Bayesian most probable field of the posterior distribution. A QL estimator of the density field - including its dark matter (DM)  component - is obtained by taking the geometric mean over the density fields of an ensemble of fully non-linear CF2-constrained simulations. The averaging process washes out the internal structure of collapsed halos, thereby filtering out the extreme non-linear virial regime and leaving behind  the QL regime.

One of the outstanding open issues of galaxy formation is that of the galaxy bias, namely the fact that the   galaxy distribution does not necessarily follow that of the DM\cite{1986ApJ...304...15B,1987Natur.326..455D}.  Studies of  the galaxy bias range from studies of the statistical relation between the distribution of galaxies and matter on large cosmological scales\cite{2012MNRAS.424..472B,2016arXiv161109787D,2017arXiv171102677S} to studies of the bias in the 
local `universe'\cite{1993ApJ...412....1D,
2002MNRAS.336.1234Z, 
2011MNRAS.413.2906D,
2015MNRAS.450..317C,
2017MNRAS.470..445N}.   
The local studies consist of the point-by-point comparison  of either the density or velocity fields within the local neighborhood. Procedures to measure  the galaxy bias depend on the nature of the data used to uncover the density field. In the case of redshift surveys a galaxy bias model needs to be assumed in order to translate from the observed galaxy density field to the underlying matter density field. Galaxies are displaced in redshift from their true positions, causing an element of circularity in the modelling and measurement of the bias that needs to be carefully addressed. Attempts to compare the velocity field constructed from redshift surveys with measured velocities\cite{2015MNRAS.450..317C,2017MNRAS.470..445N} need to further account for the tidal component of the velocity field, induced by structures outside the volume sampled by the redshift surveys.
The QL density field derived from the CF2 data   is compared here with the one  derived from the 2M++  redshift compilations\cite{2011MNRAS.416.2840L,2015MNRAS.450..317C}, and  the scale dependent non-linear galaxy bias on scales ranging upward from $ \sim 5.6 \hmpc$ is estimated. Measured velocities trace  directly the total mass density field and sample the full velocity field, hence the method does not suffer from the aforementioned difficulties involved in using redshift surveys. The downside of using velocities as constraints is the greater susceptibility  to observational biases and systematic errors compared with redshift surveys. It follows that the two approaches complement one another and both need to be used and compared.

{\bf Results}

 An ensemble of 20 constrained N-body DM only simulations has been generated in a box of side length of   $500 \hmpc$  at a resolution of $N=512^3$ (simulations and   CF2 data are described in Methods). The density and velocity fields are constructed by a clouds-in-cells algorithm on an $N=512^3$ grid and 
unless it is otherwise stated the density and velocity fields are Gaussian smoothed with a kernel of radius $R_s=2.0\hmpc$.
The averaging over the ensemble of simulations   filters out the virial highly non-linear regime and leaves behind the quasi linear (QL) density field, $\Delta^{QL}(\br)$.
(Here, $\Delta =\rho /\bar{\rho}$, where $\rho$ is the matter density and $\bar{\rho}$ is the mean density of the Universe.) 
The geometric mean taken over the density fields of the ensemble of constrained simulations is taken here to be an estimator of the QL density field,  $\Delta^{QL}$ (see Methods for justification). 
 The distribution of the Cartesian components of the smoothed velocity field is close to normal and therefore the QL velocity field is estimated by the    arithmetic mean taken over the   constrained simulations.
The scatter around the mean is calculated as well.   
The QL estimator shares many of the properties of the linear WF. Where the data is 'strong' the constructed  field is determined by the data, fairly independent of the prior model, and the scatter around the estimated field is significantly smaller than the cosmic variance.  In the limit of sparse, noisy or incomplete ("weak") data  the estimated field converges to the mean field predicted by the prior model and the scatter around the estimated field is given by the cosmic variance. Here, we choose the \LCDM\  model with  Planck cosmological parameters  (see Methods).  

The full three-dimensional structure uncovered by the QL reconstruction is explored by means of  a video and a Sketchfab interactive graphics tool\cite{2017PASP..129e8002P} (Supplementary material).
Figure 1 shows a three-dimensional visualization of the large scale structure by means of isosurfaces of the   density field. 
Figure 2 provides a detailed  description of the QL matter distribution at the equator in the Supergalactic coordinate system. 
One should recall  the general tendency of the estimated density fluctuations to diminish with distance as the quality of the data deteriorates. The statistical significance of the various structures is gauged by the signal to noise ratio, $S/N = \Delta / \sigma_\Delta$, 
where  $\Delta$ is the QL estimator of the density normalized by the cosmological mean density and $\log_{10}\sigma_\Delta$ is the  standard deviation of  the scatter in $\log_{10}\Delta$ over the ensemble of constrained simulations. At $2 \hmpc$ [$4 \hmpc$] Gaussian smoothing
for various over-densities is 
 $S/N \approx$  22 [32] for Virgo Cluster,  17 [27] for Centaurus Cluster, 9 [14] for Coma Cluster  and  9 [16]  for Perseus - Pisces supercluster (Supplementary Figures 4a and 4b). It follows that the increase in  resolution in the non-linear regime is associated with a decrease in S/N. The right hand side panel presents the  $( \Delta  -1) / \sigma_\Delta$ map and manifests the   gradually diminishing of the  statistical robustness  of the reconstruction beyond the data zone (see Methods). Yet, regions of high $S/N$ values are found at larger distances with a peak value of ~7.3 at a distance of $R=140\hmpc$ (coinciding with the Shapley Concentration) and of 10.0 at $R=160\hmpc$  (quoted $S/N$ values are for $4\hmpc$ Gaussian smoothing).


A particular redshift compilation of interest is the 2M++\cite{2011MNRAS.416.2840L} which is an augmentation of the Two MASS Redshift Survey\cite{2012ApJS..199...26H}  data with redshifts from the 
Sloan Digital Sky Survey-DR7\cite{2009ApJS..182..543A} and the Six Degree Field Galaxy Survey\cite{2009MNRAS.399..683J}.  A density field was constructed from the 2M++ survey by counting galaxies in cells, weighting them by their luminosities and further correcting for the magnitude limit by the inverse of the selection function\cite{2015MNRAS.450..317C}. The Galactic zone of avoidance  was filled by cloning data from nearby  regions. The density field was evaluated on a $256^3$ grid with a grid unit of $2.65 \hmpc$ and was Gaussian smoothed with a kernel of $4.0 \hmpc$. The raw 2M++ density field has been further re-mapped to the QL grid. The comparison between the QL and the 2M++ density fields is done at a Gaussian resolution of $4\sqrt{2} \hmpc$ (see Methods).
The selection function of the 2M++ galaxies drops drastically at a distance of $R\approx 80\hmpc$\cite{2015MNRAS.450..317C}, so beyond that distance the 2M++ density   becomes dominated by shot noise. The following comparison is therefore limited to within  $R=80\hmpc$. 
The QL and 2M++ comparison should also avoid the  zone lost to Galactic obscuration in both the 2M++ and the CF2 data - so as to focus on the most robust regions of the two data sets. This is achieved here by considering only grid points that include CF2 data points within a distance of $80 \hmpc$\cite{2016MNRAS.457..695P}. The  zone of avoidance imposed on CF2 is more extended in Galactic latitude than that imposed on 2M++ hence obscuration hardly affects our bias analysis.
Figure 3  shows the QL and 2M++   density fields at the Supergalactic equatorial plane.

The left panel of  Figure 4   shows  the probability distribution function of the tracer CF2 data points in the ($\Delta_{\rm 2M++}, \Delta_{\rm QL}$ ) plane. 
Here $\Delta_{\rm 2M++}$ is the luminosity weighted normalized  density derived  from the 2M++ redshift survey.
The distribution of the raw 2M++ densities is clearly biased with respect to the QL densities and the  bias is not linear. 
The black contours illustrate a local non-linear bias model, wherein the galaxy density at a point depends on the matter density at that point. 
We consider the simplest non-linear extension of the linear bias model and assume a power law relation of the form:
\begin{equation}
\Delta = C \Delta{^\alpha_{\rm 2M++}}
\end{equation}
The free parameters of the  model, $\alpha$ and $C$,  are found by minimizing the  variance $(\Delta_{\rm 2M++} - \Delta_{\rm CS})^2$, 
taken over all the trace data points, and are evaluated for each one of the constrained simulations, where $\Delta_{\rm CS}$ is the normalized density field of a given simulation. The mean and variance of the parameters are then taken over the ensemble of 39 simulations (20 with $N=512^3$ and 19  with $N=256^3$ all in a box of $L=500 \hmpc$). The normalization parameter $C$ is found to be $0.84\pm 0.02$, which reflects the bias in the cosmological density derived from the 2M++ survey. It is 
 renormalized here to unity so as to enforce the cosmic mean density over the resulting bias-free 2M++ density field in the full computational box. The power law index is found to be $\alpha=0.57 \pm 0.04$.
The contours of the left panel of Figure  4 show  the probability of the density of the tracer data points in the ($\Delta_{\rm 2M++}, \Delta_{\rm QL}$ ) plane, where the bias is corrected by applying Equation~1 to the raw density. Indeed, the bias is largely removed.
The right panel of Figure 4 shows the 1-point probability distribution of the  raw and the bias-corrected 2M++ densities and the QL density.  The distributions of the QL and bias-corrected 2M++ densities  are well approximated  by a lognormal distribution. The 1-point distribution of the raw 2M++ density exhibits an excess in the   low end tail of the distribution. The bias-corrected 2M++ density distribution closely  traces the QL density. 
The linear biasing factor $b$ is related to $\alpha$ as follows.
Writing $\Delta=1+\delta$, the nonlinear bias of Eq. 1 is expanded to linear order in $\delta$,
\begin{equation}
\delta \approx   \alpha \delta_{\rm 2M++}.
\end{equation}
This formulation recovers the linear bias model, which is  written as $\delta_g=b\delta$, where $\delta_g$ is  the normalized overdensity inferred from the galaxy distribution.  It follows that $b = 1/\alpha$.
The analysis of the non-linear bias has been extended to resolution ranging up to $20\hmpc$ - reaching the linear regime (Figure 5). The resolution is controlled by the Gaussian smoothing with a smoothing radius $R_s$. 
yielding   $\alpha$ ranging from  $0.58 \pm   0.04$       [$b = 1.74  \pm   0.13$] for $R_s \sim 5.6\hmpc$ 
to  $0.73  \pm  0.06$       [$b = 1.38  \pm  0.11$] at $R_s=20\hmpc$. 
The linear bias is recovered at $R_s \approx 20\hmpc$.

The present bias analysis is limited to distances less than $80\hmpc$, the region least affected by shot noise uncertainties,
 and it avoids the Galactic zone of obscuration\cite{2015MNRAS.450..317C}. 
The half mean galaxy-galaxy separation of the 2M++ survey within $80\hmpc$ is $\approx2.7 \hmpc$. Hence, a Gaussian sphere of $R \sim 5.6 \hmpc$, which has the top hat volume of radius $\sim 5.6 / 0.64 \hmpc$, contains in the mean $\approx 85$ galaxies. This corresponds to roughly $11\%$ shot noise errors within the volume in which the biasing analysis has been performed.
It follows that the 2M++ errors within that volume are much smaller than those of the QL density field and are neglected here.

There is a vast body of work aimed at determining the biasing of the galaxy distribution, most of which assume the linear bias model and  set constraints on $b$ by comparing redshift surveys with peculiar velocity 
data\cite{2002MNRAS.336.1234Z, 
2011MNRAS.413.2906D, 
2015MNRAS.450..317C,
2017MNRAS.470..445N}.  
The linear theory is invoked 
in all of these studies  to relate the density and velocity fields. Such analyses relate the linear bias factor with the cosmological density parameter so as to constrain   $\beta=f(\Omega_m) / b$ where $f(\Omega_m) \approx \Omega{^{0.55}_m}$ is the linear growth factor. The QL estimation is based on a given value of 
$\Omega_m=0.3071$. Using this value one finds $b \approx 0.52 / \beta$. 
A comparison of velocity linearly inferred from the same 2M++ density field used here and the SFI++ data of velocities has found $\beta=0.43 \pm 0.02$\cite{2015MNRAS.450..317C}, which yields $1/b =  0.83 \pm   0.04$.
The most recent determination of the biased relation between the 2MRS density field and the Cosmicflows-3 velocities finds\cite{2017MNRAS.470..445N} 
$  0.34  <  \beta  <  0.52$, which corresponds to $  0.65 <  1 / b  <      0.99$.  
This agrees with the current work (at $R_s=20\hmpc$), 
 $0.73  \pm  0.06$.

A qualitative  support for our $\alpha=0.57 \pm 0.04$  bias model (Equation 1) is provided by  a recent study of the  SDSS redshift survey\cite{2017arXiv170709002W}, in which the underlying matter density is estimated by means of constrained simulations - a study akin in its methodology to the present one. The authors of the study have not quantified the bias $\Delta$ vs $\Delta_g$ relation but have presented it in Fig. 2 of that study. Estimating the relation from the figure, over the density range exhibited by the QL density field,   we obtain $\alpha\sim0.53$. A direct comparison is hindered because the estimation of the galaxy number density field is different from the one used here but this quantitative support is encouraging.

The non-linear bias relation is further tested by the comparison of the raw and the bias-corrected 2M++ density field with the QL density in Local Group-centric spheres of varying radius (Figure 6). The raw 2M++ Local Group-centric  density profile is markedly different to that of the QL one, but the bias correction brings it to a close agreement. 

The QL density and velocity fields are rich with structures and many aspects deserve a detailed study and analysis. Here we focus on the density and radial velocity profile of the structure outside the collapsed core of the Virgo Cluster. The cluster center is identified with the peak of the QL density field   at ${\bf R}{^{QL}_{Virgo}}=[-4.9,     12.7,      1.0] \hmpc$, - a displacement of $2.8 \hmpc$ from the galaxy M87, the centroid of the optical galaxies in the cluster (at $[-2.70, 11.70, -0.52]\hmpc$). Figure 7 shows the Virgocentric density  and radial velocity profiles.  The former records the mean density within spheres centered on Virgo while the latter gives the physical (peculiar plus the Hubble expansion) radial velocity in Virgo-centric shells. The profiles and their scatter are taken over the raw simulations with no further smoothing. 
Of particular interest is the Virgo Cluster turn around (TA) radius and the mass enclosed within it. We find here   $R_{TA}= 6.7 \pm 1.5  \hmpc$ and $M_{TA}= (5.3  \pm 1.6 )\times 10^{14} \hmsun$.  The resolution of the simulations used here   barely enables the determination of the turn around radius and mass of Virgo, yet the current results are    close to other determinations of these measures of Virgo. 
Table 1 presents  the mass and radius of the present paper and   four other studies. 
1. Current study - the TA mass and radius measured with respect to the Virgo cluster associated with the local density maximum of the QL density field. 
2. A study of the Virgo cluster by means of simulations constrained by the CF2 data that recently reported\cite{2016MNRAS.460.2015S} the TA mass and radius of the Virgo cluster. That study differs from the one reported here in two ways: (a) The bias correction used to undo the Malmquist bias of the CF2 data in that paper differs from the one used here. (b) The density and radial velocity profiles, used to calculate the TA mass and radius were calculated there with respect to the DM halo that was chosen as a proxy to the actual Virgo. Here  it is calculated with respect to local density maximum of the QL density field. 
3. A reconstruction of the orbits and 3D velocities of all nearby (within $\approx 30\hmpc$) galaxies by the numerical action method\cite{2017ApJ...850..207S} constrained by Cosmicflows-3 distances\cite{2016AJ....152...50T}. 
4. A model fitting of the  Virgocentric  velocity-distance relation\cite{1992A&A...260...17T} that was used to estimate the mass and radius 
of the zero-velocity surface\cite{2006NewA...11..325P}.   
5. A recent study based on tip of the red giant branch observations with Hubble Space Telescope that estimated the Virgo TA radius from the distances and velocities of galaxies to the foreground of the Virgo cluster\cite{2014ApJ...782....4K}. Assuming the spherical symmetric infall model and the standard model of cosmology the TA mass is inferred. All numbers in Table 1 have been converted to the $h^{-1}$ scaling.  In considering the present estimation of the Virgo TA radius one needs to recall that the CIC grid spatial resolution is $2\hmpc$. Given that, the agreement  with the other estimations of the Virgo TA mass and radius is reassuring - it suggests that the TA parameters of the Virgo cluster are robustly determined and it lends further support to the present QL reconstruction.

  \bigskip
{\bf Discussion}\newline 
The QL analysis of peculiar velocities enables  the reconstruction of the  underlying matter density and velocity fields out to distances exceeding $100 \hmpc$ based on the CF2 galaxy distance and radial velocity data only and making no other assumption but that of the \LCDM\ cosmology. 
This opens a new window into the dark sector of our local patch of the Universe, namely the distribution of the DM around us.
Peculiar velocities are the only means in our astronomical toolbox by which the local DM distribution can be mapped. 
The reconstruction from velocities avoids the two shortcomings inherent to the reconstruction from redshift surveys. 
These are the need to assume: a.   a bias model to relate the observed distribution of galaxies to the underlying mass distribution;  b. the contribution of structures beyond the computational box and in the zone of obscuration in order to construct the full velocity field.  The use of velocities, on the other hand, faces its challenges in deriving a dataset of constraints from   the raw astronomical data. Two such difficulties to overcome are the correction of the Malmquist-like biases and the grouping of the data.

In regions where the QL reconstruction is dominated by the data, namely regions densely sampled by accurate peculiar velocities, the turnaround radius and mass of rich clusters can be robustly estimated. The turnaround   mass is the cleanest and most physical estimator of the mass of a cluster and indeed  it has been applied here to the Virgo cluster. With the availability of next generation Cosmicflows-3 data the turnaround mass of more distant clusters will be assessed.

Galaxies are formed 'biased' - their distribution does not necessarily follow that of the DM. 
The comparison of the luminosity weighted galaxy density field of the 2M++ redshift compilations  with the QL matter distribution provides a window into bias on the resolution range of $5.6 \leq R_s \leq 20 \hmpc$. A non-linear local bias model with a power law index $\alpha$ has been assumed here, with $\alpha$ ranging from  $0.58 \pm   0.04$       [$b = 1.74  \pm   0.13$] for $R_s \sim 5.6\hmpc$ 
to  $0.73  \pm  0.06$       [$b = 1.38  \pm  0.11$] at $R_s=20\hmpc$ ($b=1/\alpha$).
The estimated $\alpha$ is in agreement with earlier studies performed in the linear regime.  
Galaxies are more 'biased' in the quasi-linear regime compared to the linear one.  Our finding of  the non-linear bias at the $R_s \sim 5.6 \hmpc$  scale  is qualitatively supported by the recent analysis of the SDSS redshift survey by the ELUCID collaboration\cite{2017arXiv170709002W}. (Figure 2 of the original version posted on the arXiv. The figure was removed from final version because of editorial considerations - private communications with the authors).  Galaxy formation is an inherently non-linear process and the appreciation  of the bias of galaxies is crucial to the understanding their formation and evolution. The QL analysis brings down the scale on which the bias is studied - thereby putting stronger constraints on models of galaxy formation.

\newpage
\begin{methods} 

{\bf Data, Bayesian framework  and prior model:}\newline
{\bf Data:}
The present study is based on the CF2 dataset\cite{2013AJ....146...86T},   that extends sparsely to distances of $\sim300\hmpc$ (redshift $z\approx0.1$). It consists of  8,161 entries with high density of coverage inside $\sim100\hmpc$, a region called here the data zone. 
A grouped version of the Cosmicflows-2 data is used here, in which all galaxies forming a group, of two or more, are merged to one data entry. The grouped CF2 dataset consists of 4,885 entries.  The grouping 
is   an effective way of filtering out the internal virial motion and recovering the motion of the group as a whole.
This is an effective linearization of the data.
\newline
{\bf Bayesian framework:} 
The linear  WF/CRs\cite{1995ApJ...449..446Z} and their quasi-linear  QL  extension are applied within a Bayesian framework that rests on two pillars: the constraining  observational  data and the assumed prior model. The Universe introduces itself into the reconstruction of the local `universe' by means of the  CF2  data of galaxy   velocities. 
Structure is assumed to evolve via gravitational instability from a primordial Gaussian perturbation field, whose properties are defined by the assumed power spectrum. 
\newline
{\bf Prior model:} The prior model is assumed to be the standard $\Lambda$ Cold Dark Matter (\LCDM) model of cosmology, with parameters determined mostly by the Planck analysis of the Cosmic Microwave Background (CMB) radiation\cite{2014A&A...571A..16P}. 
Where the data is absent or is   not robust enough the prior model dominates the construction. The statistical consistency of the CF2 data with the prior \LCDM\  
model\cite{2016MNRAS.461.4176H} lends support to the robustness of the reconstruction of the large scale structure presented here. However we are aware of the tension between the local determination of Hubble's  constant ($H_0$) and its   derived values   based on an adopted cosmological model and measurements of the CMB\cite{2017NatAs...1E.121F}. The value of $H_0$ consistent with the measured distances and redshifts of the CF2 data is $H_0=75 \pm 2$ \kmsMpc\  \cite{2013AJ....146...86T, 2016AJ....152...50T}.
Yet, the power spectrum of the underlying primordial perturbation field is evaluated with the Planck derived value of $H_0=67$\ \kmsMpc\ \cite{2014A&A...571A..16P}. Distances  are expressed here in units of $\hmpc$ where $h=H_0 /100$\ \kmsMpc.

{\bf Constrained N-body simulations  and the QL estimator:} \newline
{\bf Constrained simulations:} 
The WF/CRs methodology applied to the  CF2 data augmented by the reverse Zeldovich approximation (RZA) algorithm\cite{2013MNRAS.430..902D,2013MNRAS.430..888D,2013MNRAS.430..912D}  are used to construct constrained initial conditions for the numerical simulations. 
These have been run using the \texttt{Gadget2} \emph{N}-body code\cite{2005MNRAS.364.1105S}, which implements 
a Tree-PM gravity solver.
The runs have been performed in a DM-only fashion, with $512^3$ particles  in a periodic box of $500 \hmpc$
and a co-moving softening length of $25 \hkpc$.
 An ensemble of 20 constrained simulations has been constructed. Another ensemble of 19 similar simulations but with $256^3$ particles was constructed and used to study the non-linear bias parameters. 
A clouds-in-cells (CIC) algorithm has been used to construct the 
density and velocity fields on a $512^3$ grid from the particles distribution. The application  of the CIC algorithm to the construction of the velocity field is adaptive. Where empty cells are found the cells are doubled in length (in each Cartesian direction) and the velocity is calculated on the coarser grid. The procedure is repeated until a non-empty cell is found. 
\newline
{\bf Tests of the  constrained simulations:} 
The WF/CRs/RZA methodology for generating constrained initial conditions and thereby constrained simulations was throughly tested\cite{2013MNRAS.430..902D,2013MNRAS.430..888D,2013MNRAS.430..912D}.  The test consists of running a (peculiar velocities) constrained simulation in a BOX160 ($160\hmpc$ on its side). Such a simulation reproduces the main structural features of the local universe (Virgo and Coma clusters, Local Void, Local Supercluster,  Great Attractor and the Perseus - Pisces supercluster). That simulation was ÔobservedÕ in a manner that emulates the CF2 data, including the placing of a mock observer in a LG-like object and a CF2-like mock catalog was created. Ten constrained simulations, based on the WF/CRs/RZA methodology and an assumed \LCDM\ cosmology, were run. The bottom-right  panel of Fig. 3 of Doumler et al\cite{2013MNRAS.430..912D} presents a cell-by-cell comparison of the $5.0\hmpc$ Gaussian smoothed density field of the target simulation and the constrained ones. The figure shows an unbiassed scatter with an r.m.s. scatter of 0.58 (in $\Delta$). It is noted here that the CF2-like mock data was constructed by placing the mock data points at their exact positions and all uncertainties were assigned to the mock observed velocities. It follows that, to the extent that the Malmquist-like biases are corrected, the current algorithm of running constrained simulations in the \LCDM\ cosmology from Cosmicflows-like data results in an unbiased sampling of the underlying density field.  
\newline
{\bf Cosmological parameters:}
The N-body simulations are run with the following cosmological parameters: 
$\Omega_M = 0.31$ (matter cosmological density parameter),
$\Omega_\Lambda = 0.69$ (dark energy cosmological density parameter),
$\Omega_b = 0$ (baryon cosmological density parameter),
$H_0 = 0.67\ km\ s^{-1} Mpc^{-1}$ (Hubble's constant),
$\sigma_8 = 0.83$ (normalization of power spectrum),
$n_s = 0.968$ (power spectrum spectral index) and 
$z_{init} = 80$ (initial redshift of the simulations). 
\newline
{\bf Smoothing:} Throughout the paper smoothing is performed by means of a Gaussian kernel. Unless it is otherwise stated,  density and velocity fields are smoothed with a Gaussian kernel of width of 2.0$\hmpc$.
The Virgocentric density and velocity profiles are calculated from the  CIC grid with no further smoothing.
The original 2M++ density field is smoothed with a kernel of 4.0$\hmpc$. It has been re-mapped to the grid of the QL construction (500$\hmpc$ and $512^3$ cells). The ruggedness of the mapping has been smoothed over by a further 4.0$\hmpc$ Gaussian smoothing, resulting in an effective $4\sqrt{2} \hmpc$ Gaussian smoothing. The comparison of the 2M++ and the QL density fields is done under this effective smoothing. 
\newline
{\bf QL estimator:} The non-linear density field outside the virial regions of collapsed halos, namely the QL regime, is lognormal distributed\cite{1991MNRAS.248....1C}. The density  fields of the constrained simulations properly sample the density field of the local `universe' and therefore their 1-point distribution is close to lognormal. It follows that the geometric mean field is close to the median of the constrained density fields and it is taken here to be the estimator of the QL density field. Yet, a price has been paid here. The  mean of $\Delta^{QL}$ taken over the computational box deviates from unity - it equals to 0.81, compared with the mean taken over the individual realizations where the mean is unity by construction. The quantitative  analysis involving the calculation of errors is performed with respect to the ensemble of realizations. The qualitative analysis of the reconstructed large scale structure is based on the geometric mean estimator.

{\bf Comparison of the CF2 QL and the  2M++ density fields:} \newline
Figure 3  shows the QL, the raw 2M++ and the bias-corrected 2M++ ($\log_{10}$ of the) density fields evaluated at the Supergalactic equatorial plane. The bias correction significantly reduces the residual of the 2M++ from the QL density fields.
The figure further shows the logarithmic difference between the QL and the bias-corrected density field. The map of the residual does not appear to be correlated with either the QL or the bias-corrected density fields.

{\bf Comparison with the linear Wiener Filter construction:} \newline
The QL construction adds small scale power to the linear WF construction. This effect is clearly illustrated in Supplementary Figures 1 and 2 . The former compares the dimensionless divergence of the velocity field, $-\nabla\cdot\bv/H_0$, of the QL density field with the linear WF fractional over density ($\delta$) field. In the linear theory, $-\nabla\cdot\bv$ scales with $\delta$. The latter figure compares the QL and  the linear WF velocity field using a graphical visualization consisting of streamlines\cite{2017NatAs...1E..36H}. Both comparisons show that the QL dynamics preserves the large scale structure while refining the structure on smaller, non-linear, scales.

{\bf QL density field: mean and scatter} \newline
Supplementary Figure 3 visualizes the statistical robustness of the QL construction of the density field. The figure shows the 'signal', namely the QL density field, the 'noise', i.e. the scatter among the different realizations, and the signal-to-noise ratio, S/N. The  S/N is depicted by showing the deviation of the density from its mean, $\delta = \Delta-1$. The 'signal' and 'noise' are evaluated at the supergalactic equator.
The three-dimensional QL density and the corresponding S/N ratio are visualized in Supplementary Figures 4a and 4b by a sequence of supergalactic Aitoff projections in shells of radii 
$R=10, 20, ..., 80 \hmpc$.  The right panels of the figure present the S/N estimator, where here the signal corresponds to the full density, $\Delta$, in order to to emphasize the appearance of the overdense regions.

\end{methods}

\newpage
\bibliography{biblio_QL}

\begin{addendum}
 \item[Correspondence] Correspondence and requests for materials
should be addressed to Y.H.~(email: hoffman@huji.ac.il).
 \item The help provided by Guilhem Lavaux in using the 2M++ density field is highly appreciated. Adi Nusser is gratefully acknowledged for his careful reading of the paper and his critical remarks.
 Support has been provided by the Israel Science Foundation (1013/12), the Institut Universitaire de France, the US National Science Foundation, Space Telescope Science Institute for observations with Hubble Space Telescope, the Jet Propulsion Lab for observations with Spitzer Space Telescope and NASA for analysis of data from the Wide-field Infrared Survey Explorer. 
  JS acknowledges support from the Astronomy ESFRI and Research Infrastructure Cluster ASTERICS project, funded by the European Commission under the Horizon 2020 Programme (GA 653477), as well as from the Òl$'$Or\'eal-UNESCO Pour les femmes et la ScienceÓ and the Centre National d$'$\'etudes spatiales (CNES) postdoctoral fellowship programs.
 GY thanks MINECO/FEDER   (Spain) for financial support under project grant AYA2015-63810-P. We  thank the Red Espa\~nola de  Supercomputaci\'on  for granting us computing time in the Marenostrum Supercomputer at the BSC-CNS where the simulations used for  this paper have been performed.
 \item[Competing Interests] The authors declare that they have no
competing financial interests.

\end{addendum}

\newpage
{\bf Figure legends}

\bf{Figure 1}: Three-dimensional visualization of the density field by means of isosurfaces. The surface shown in grey is associated with a density of $\Delta=1.2$, while surfaces shown in nuances of red are associated with higher  values of $\Delta=1.7, 2, 2.3, 2.7, 3.$ The red, green, blue $50\hmpc$-long arrows materialize the SGX, SGY, SGZ axes of the supergalactic coordinate system. 

\bf{Figure 2}: 
Contour maps   of the QL density field and of its statistical uncertainty. 
Left panel: The    $\log_{10}\Delta^{QL}$ field (left panel) of a slice on the supergalactic equatorial plane. The solid thick line corresponds to the mean cosmological density, blue contour lines correspond to underdense regions, and the black lines (with the gray scale color map) correspond to overdense regions. The contour spacing of $\log_{10}\Delta^{QL}$  is $0.2$.
Prominent objects cut by  the supergalactic equator are the cluster in Virgo (at [SGX,SGY] $\approx[0,10] \hmpc$), Centaurus ($\approx[-40,20] \hmpc$) and Coma ($\approx[-10, 70] \hmpc$), the Shapley concentration ($\approx[-(100\ $--$\ 150), (50\ $--$\ 100)]\hmpc$) and   Perseus - Pisces supercluster ($\approx[(25\  $--$\  50), -(40\ $--$\ 10)] \hmpc$). 
Right panel: the signal-to-noise   contour map shows the   ratio of the mean to the scatter, $(\Delta^{QL} - 1)/\sigma_\Delta$, with contour spacing 2.0.

\bf{Figure 3}:  
Comparison of the QL (upper-left), the 2M++ raw (upper-right) and the 2M++  bias-free (bottom-left)   density   fields. 
All density fields are Gaussian smoothed with $R_g \sim 5.6  \hmpc$. The $\log_{10}$ of the density field is shown with contour spacing of 0.1.
The bottom-right panel shows the $\log_{10}$ of the ratio of the QL and bias-free 2M++ density fields (contour spacing of 0.1).

\bf{Figure  4}: 
Left. A comparison of the   raw  and the bias-free  2M++ with the QL density fields evaluated at the positions of the CF2 data points, used as tracers of the density field.
The joint probability of the density is evaluated at the tracer points in the 2M++ and QL densities. Blue-scale color map shows the probability of the raw 2M++ data and black contours give the probability of the bias-free 2M++ density field. The raw 2M++ is clearly biased with respect to the QL density. 
Right. The probability distribution functions of the QL (solid black), the raw (dot-dashed red) and the bias-free (dashed blue) 2M++ density fields.

\bf{Figure  5}:
The bias power index $\alpha$ is plotted against the Gaussian smoothing length, $R_s$. The mean and standard deviation of $\alpha$ are calculated over the ensemble of constrained simulations. The scale dependence of $\alpha$ exhibits  a diminish in the biasing (i.e. $\alpha$ tends toward unity) in the transition from the QL scale of $R_s \approx 5$ to the linear scales ($R_s {^>_\sim} 10 \hmpc$). The equivalent  $\alpha = 1/b$ of the density-velocity linear analysis is shown by the two  blue data points with arbitrary assigned resolution of $R_s=21$\cite{2015MNRAS.450..317C} and $22\hmpc$\cite{2017MNRAS.470..445N}. The effective window functions of the linear density-velocity comparisons are very different from the Gaussian window function used here and therefore the  resolutions scales are  introduced  for the sake of presentation only.

\bf{Figure  6:} 
The  spherical mean density as a function of depth R for the QL (solid line, black), the raw 2M++ (dot-dashed, red) and the bias-free 2M++ (dashed, blue) density fields. 
The QL density field normalized such that its mean value within the computational box equals its cosmological mean value (see methods).

\bf{Figure  7}: 
Left. Virgocentric density and velocity profiles:
Curves represent the median (solid black line) and the 25\% and 75\% percentiles (dashed blue lines) of the normalized density within spheres of radius R centered on the Virgo cluster (at $[-4.9,     12.7,      1.0]\hmpc$). The median and the percentiles are calculated over the ensemble of simulations. 
Right. The mean radial (peculiar + Hubble flow) velocity within shells of radius R ($\pm 1\hmpc$) is shown, with error bars that correspond to the standard deviation calculated over the ensemble of simulations. The dashed red line shows the unperturbed Hubble expansion for reference.

\newpage

{\bf  Table 1. Virgo cluster: turn-around mass and   radius}\newline\begin{tabular}{lccc}
\hline
                                                                                                & $M$ [$10^{14}\hmsun$] & $R_{TA}$ [$\hmpc$]\\
\hline
CF2 QL density     (present paper)                                                          & $5.3  \pm 1.6$ & $6.7 \pm 1.5$ \\
CF2 dark matter halos based method\cite{2016MNRAS.460.2015S}    & $5.0 \pm 0.7$  & $2.0 \pm 0.4$\\
CF3 NAM method\cite{2016AJ....152...50T}                                           & $6.2 \pm 0.2$  & $5.5 \pm 0.2$ \\
Virgocentric model fitting\cite{2006NewA...11..325P}                             & $7.7 \pm 0.8$  & $6.0 \pm 0.5$ \\
Infall of galaxies onto the cluster (HST data)\cite{2014ApJ...782....4K}  & $5.8 \pm 1.7$ & $5.2 \pm 0.5$ \\
\hline               
\end{tabular}

\newpage

\begin{figure}
\label{fig1a}
\includegraphics[width=1.\textwidth,angle=-00]{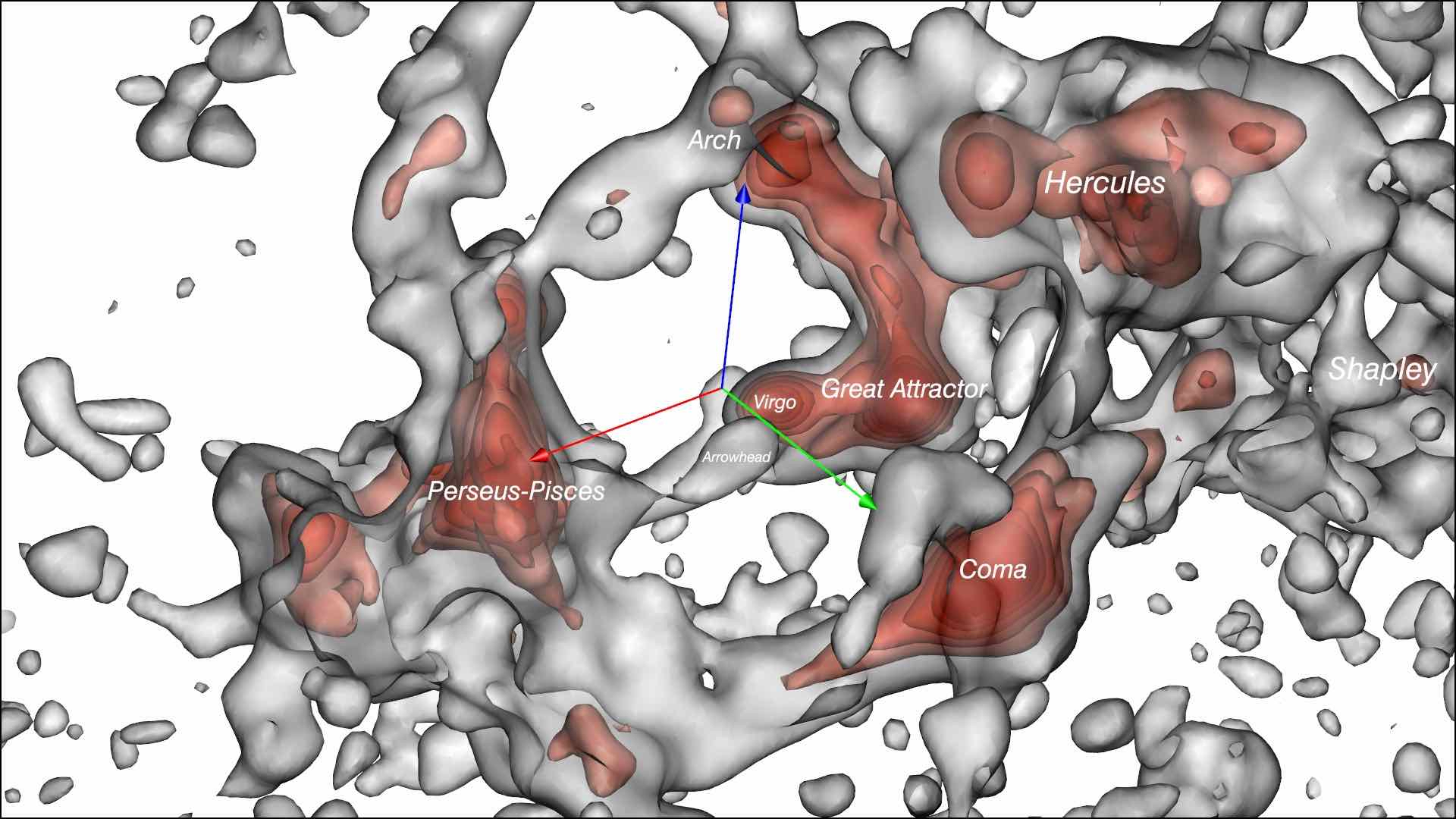}
\bf{Figure 1
}
\end{figure}


\begin{figure}
\label{fig:S2N}
\hskip -1.0cm
\includegraphics[width=.57\textwidth,angle=-00]{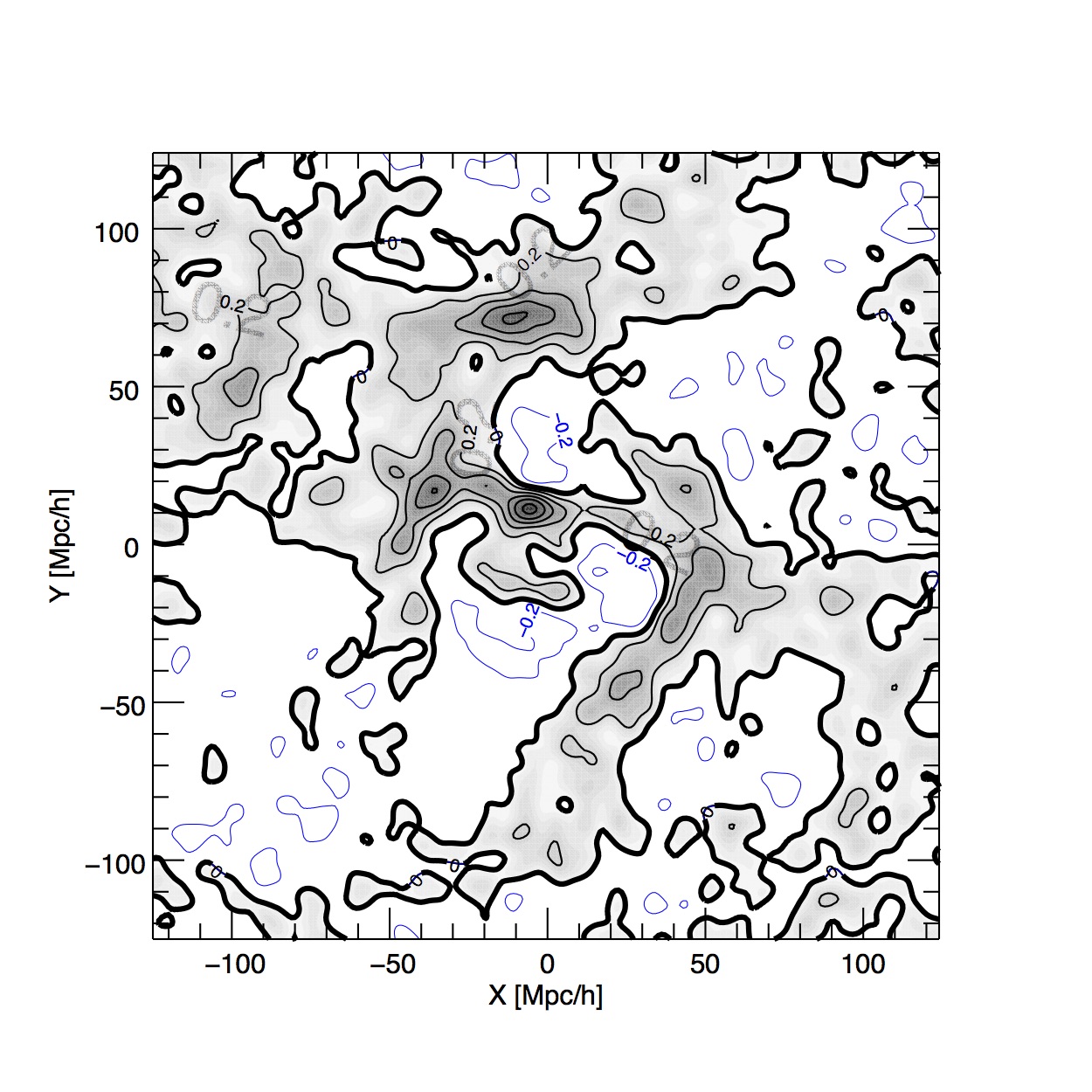}
\hskip -1.0cm
\includegraphics[width=.57\textwidth,angle=-00]{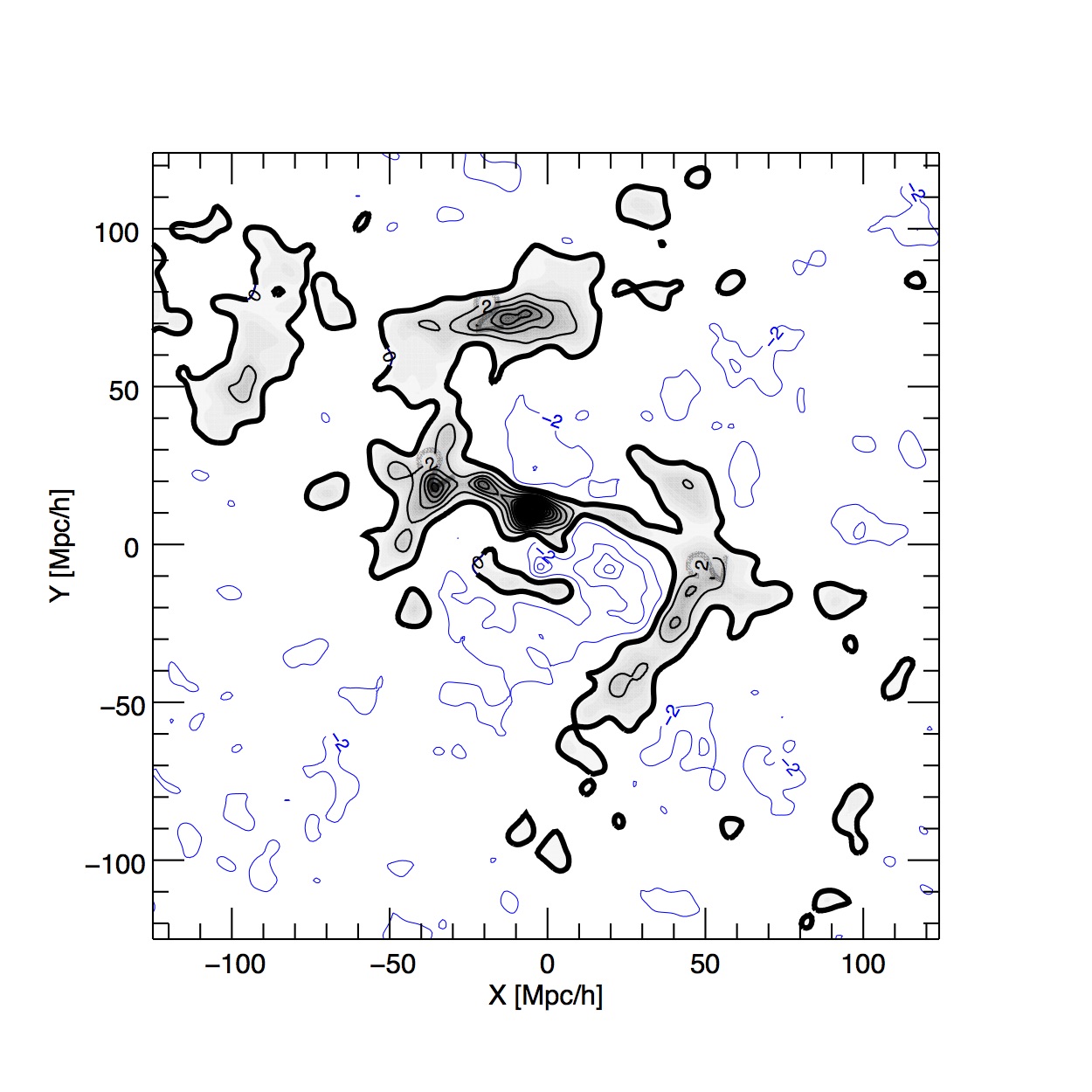}
\newline
\bf{Figure 2 
}
\end{figure}

\begin{figure}
\label{fig:QL-bias-free-TMRS}
\includegraphics[width=.5\textwidth,angle=-00]{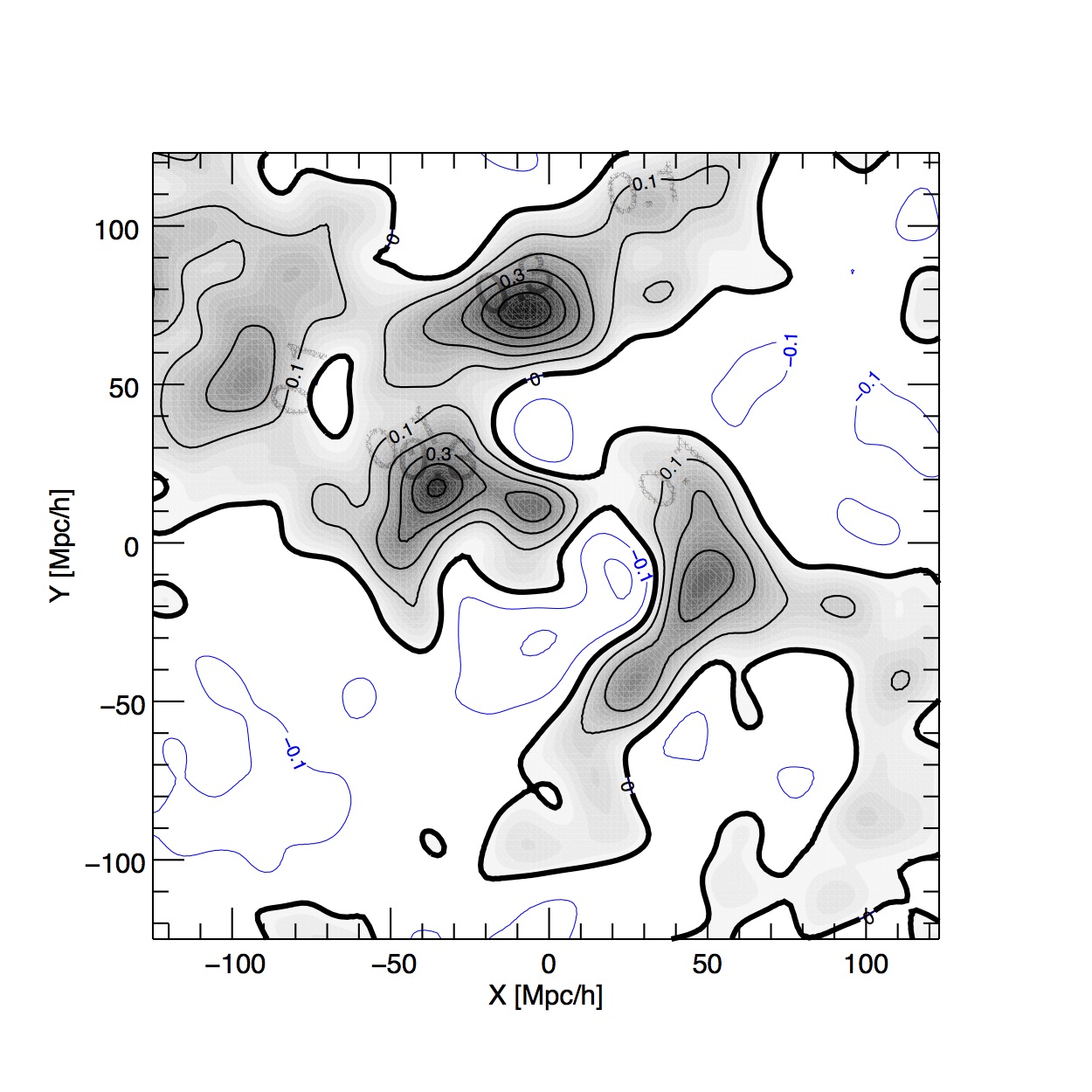}
\includegraphics[width=.5\textwidth,angle=-00]{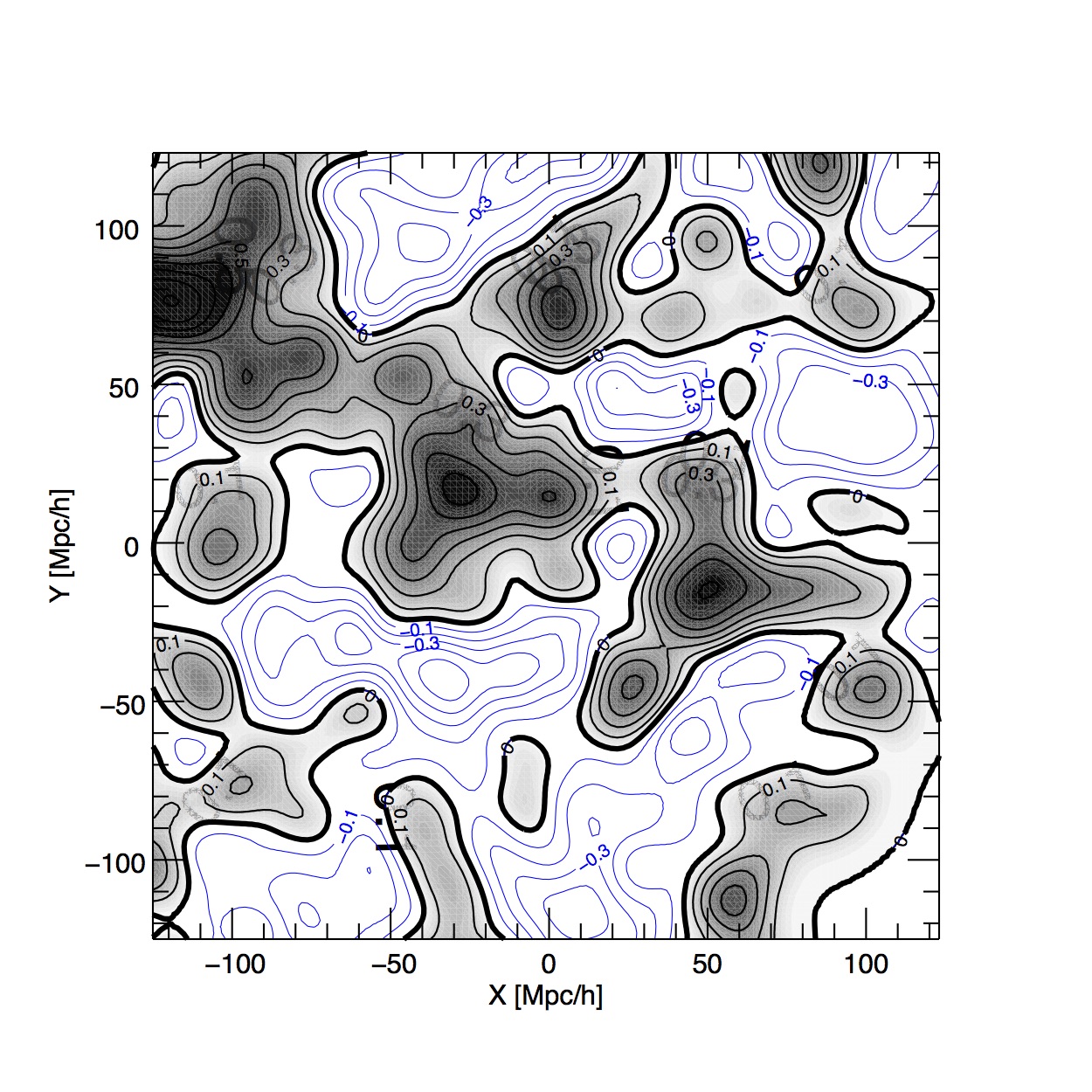}
\includegraphics[width=.5\textwidth,angle=-00]{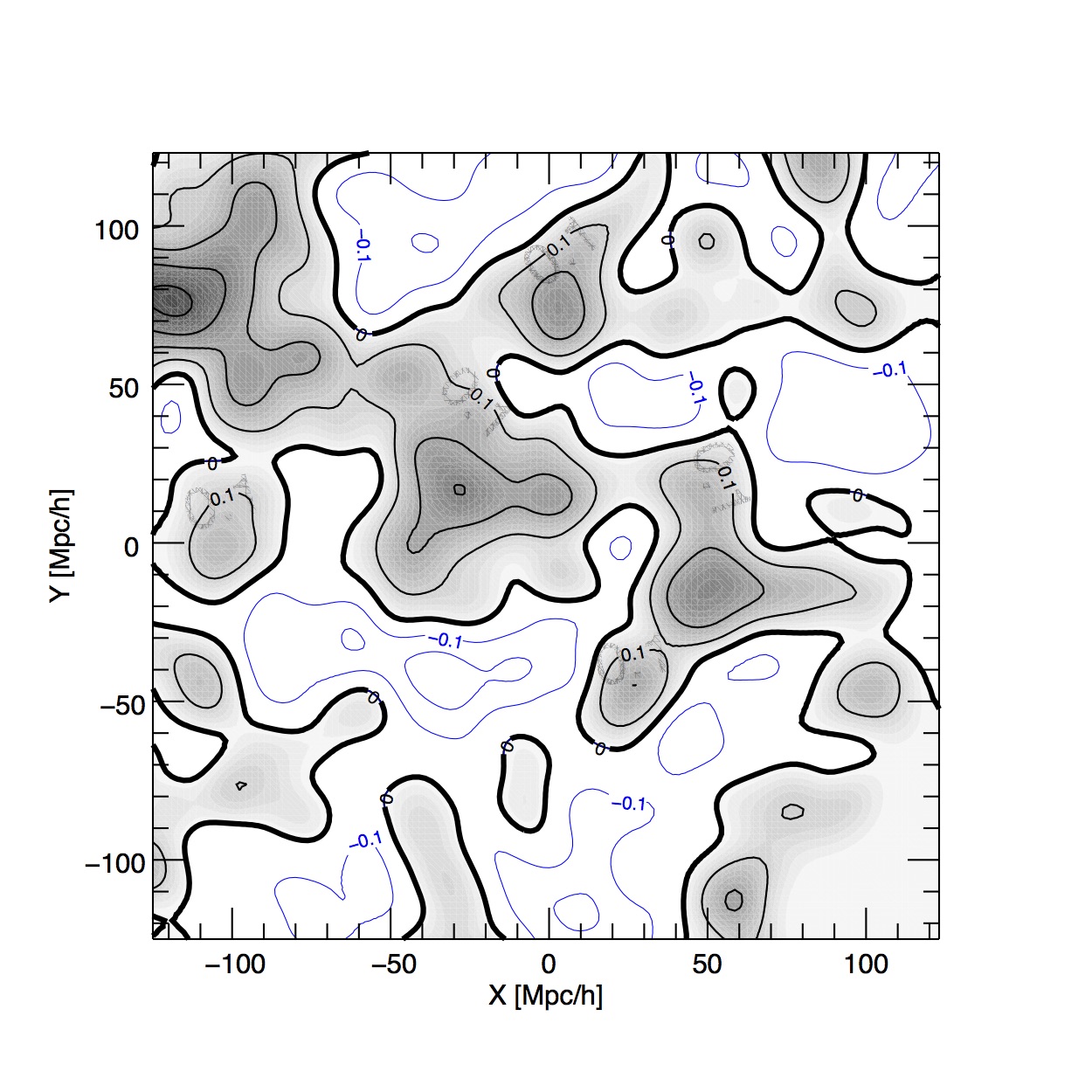}
\includegraphics[width=.5\textwidth,angle=-00]{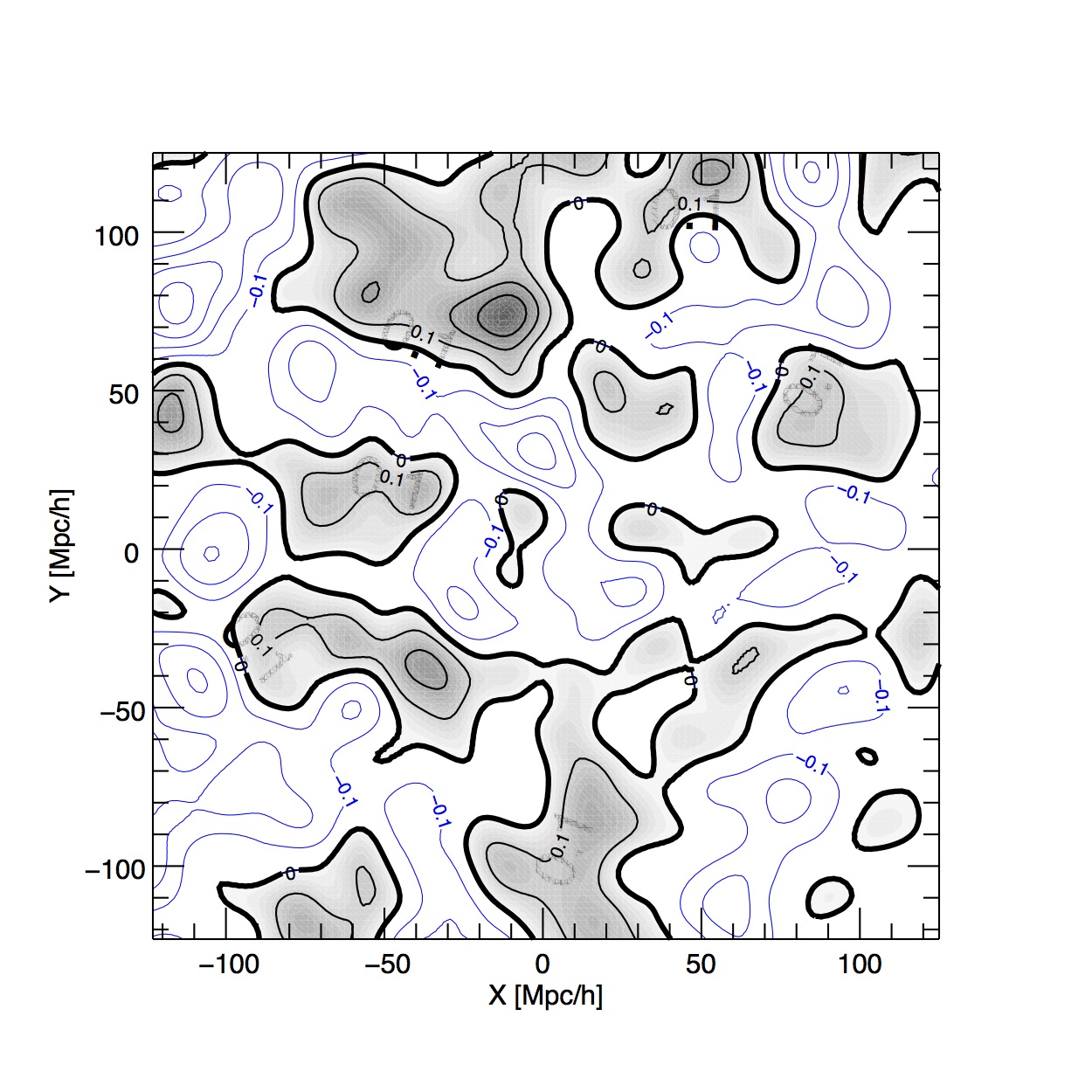}
\newline 
\bf{Figure 3  
}
\end{figure}

\begin{figure}
\label{fig5}
\hskip -0.35cm
\includegraphics[width=0.47\textwidth,angle=-00]{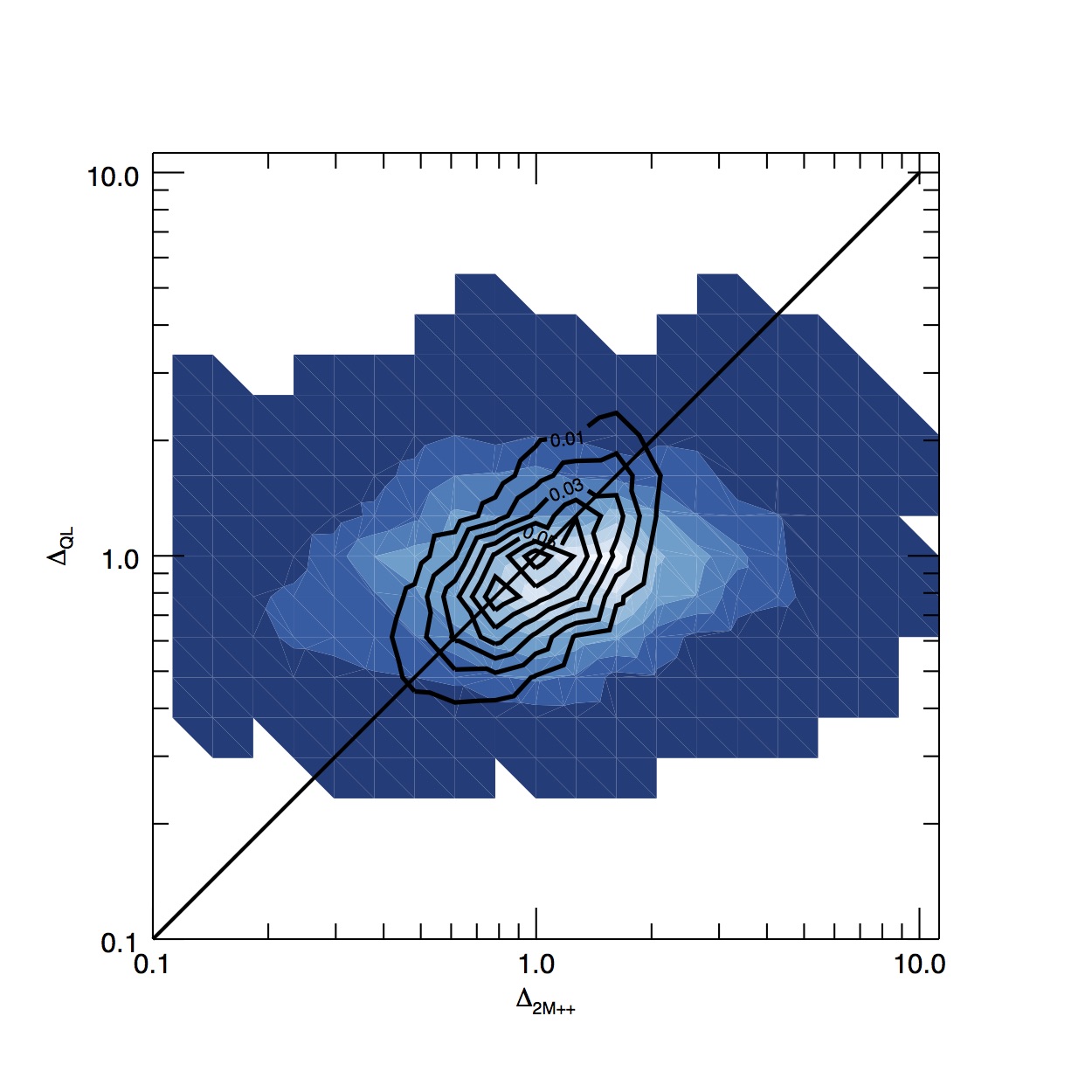}  
\hskip -0.3cm
\includegraphics[width=0.57\textwidth,angle=-00]{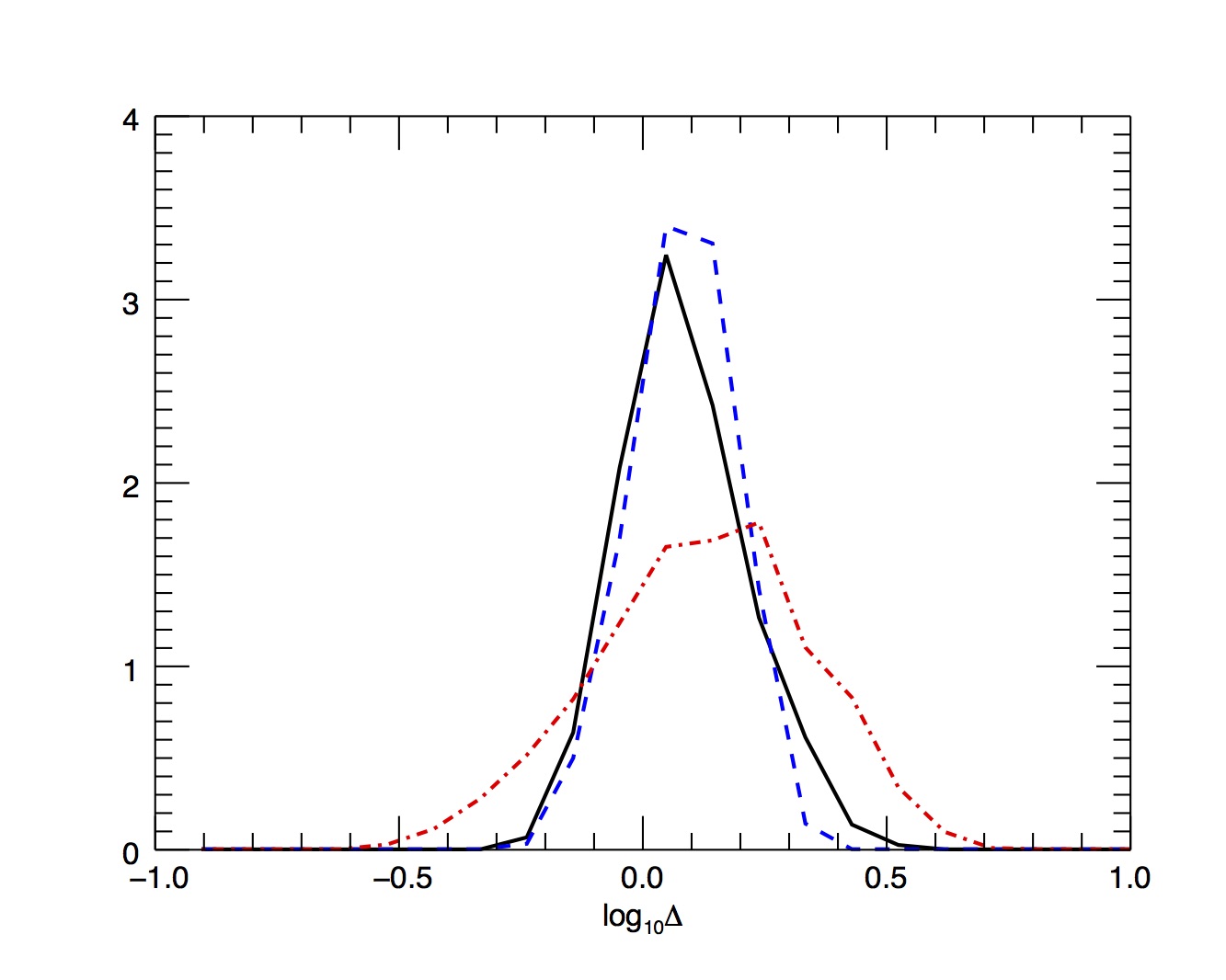}
\newline
\bf{Figure  4
}
\end{figure}

\begin{figure}
\label{fig6}
\includegraphics[width=1.0\textwidth,angle=-00]{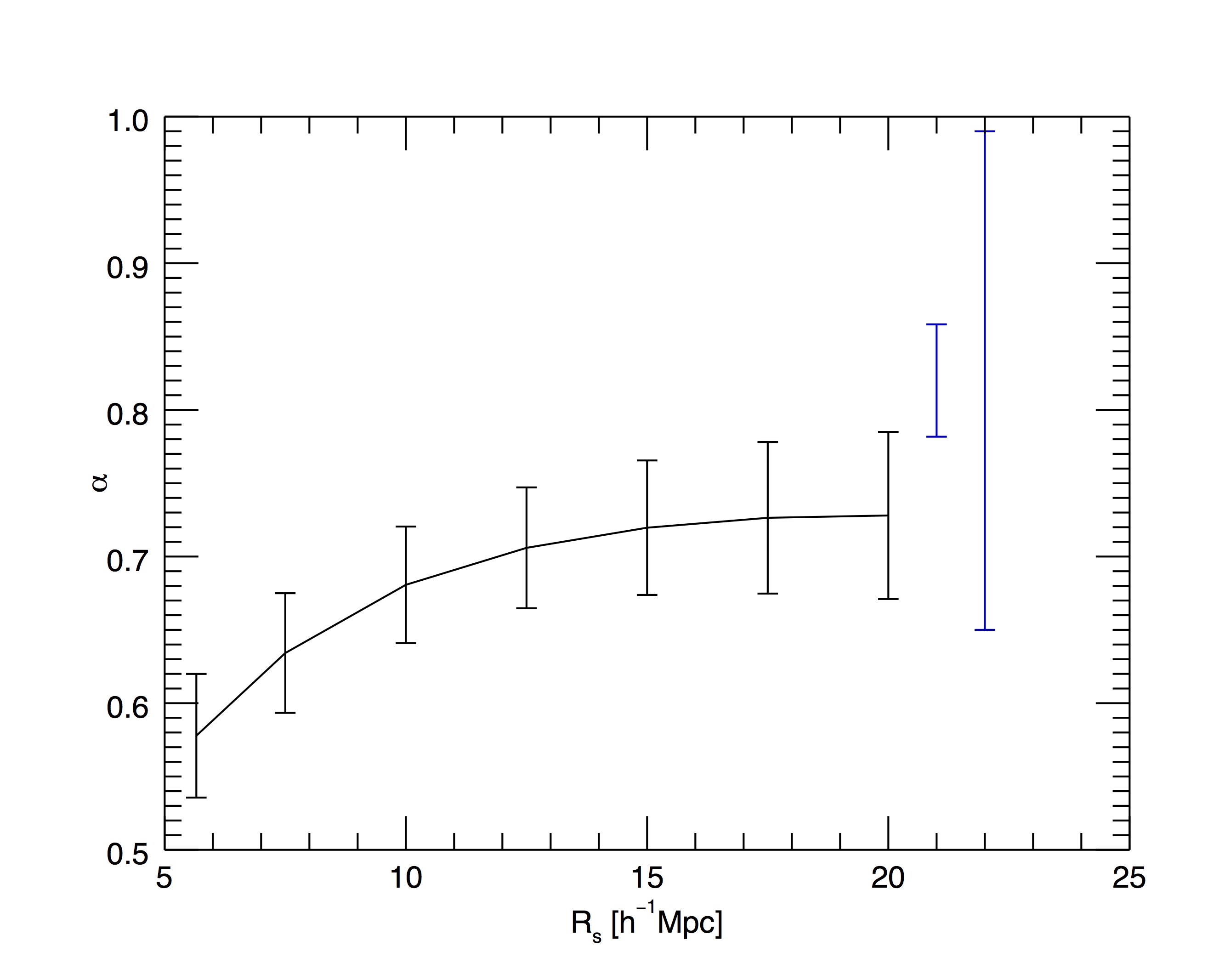}  
\newline 
\bf{Figure  5 
}
\end{figure}

\begin{figure}
\label{fig7}
\includegraphics[width=1.0\textwidth,angle=-00]{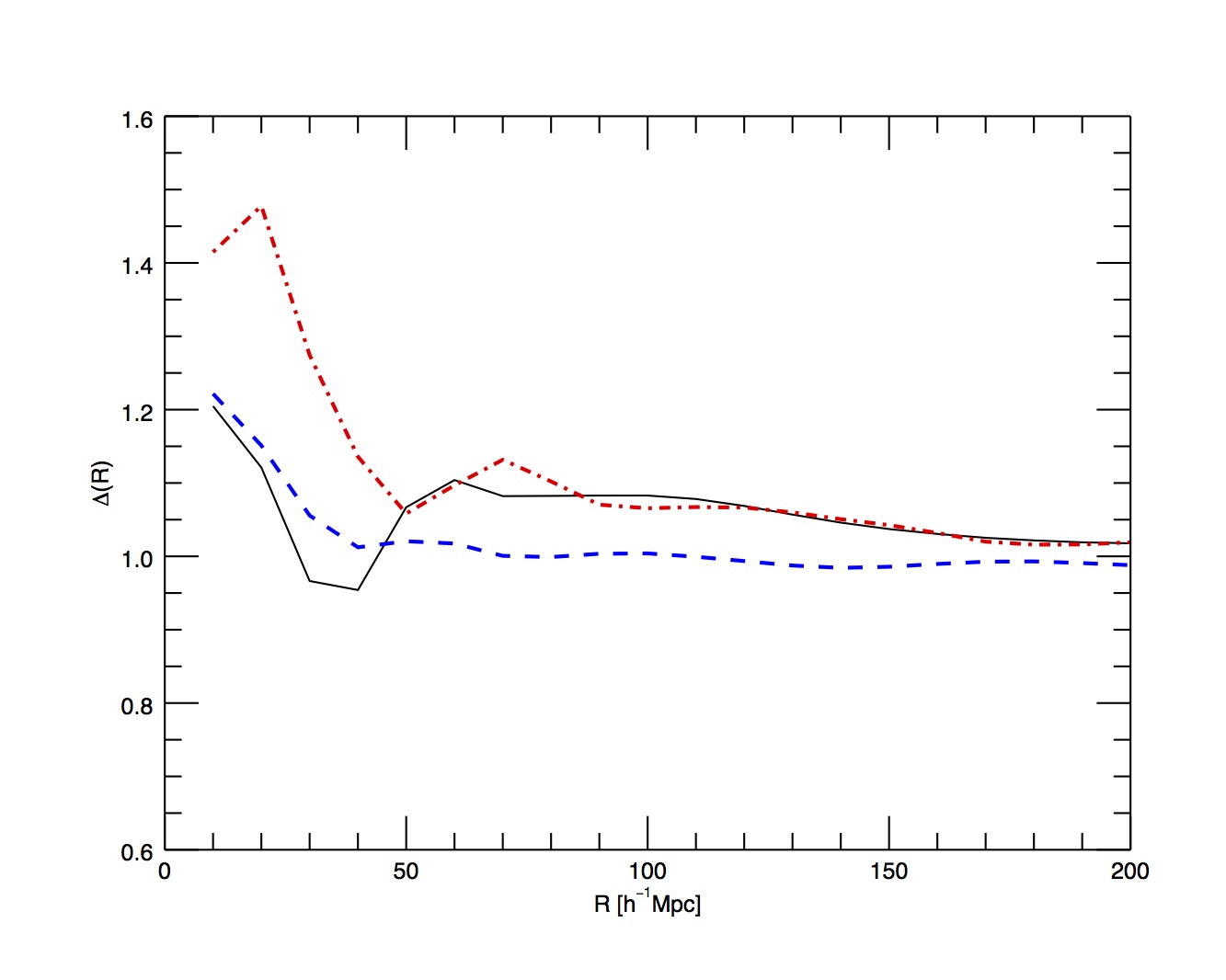}  
\newline 
\bf{Figure  6 
}
\end{figure}

\begin{figure}
\label{fig8}
\hskip -0.35cm
\includegraphics[width=0.5\textwidth,angle=-00]{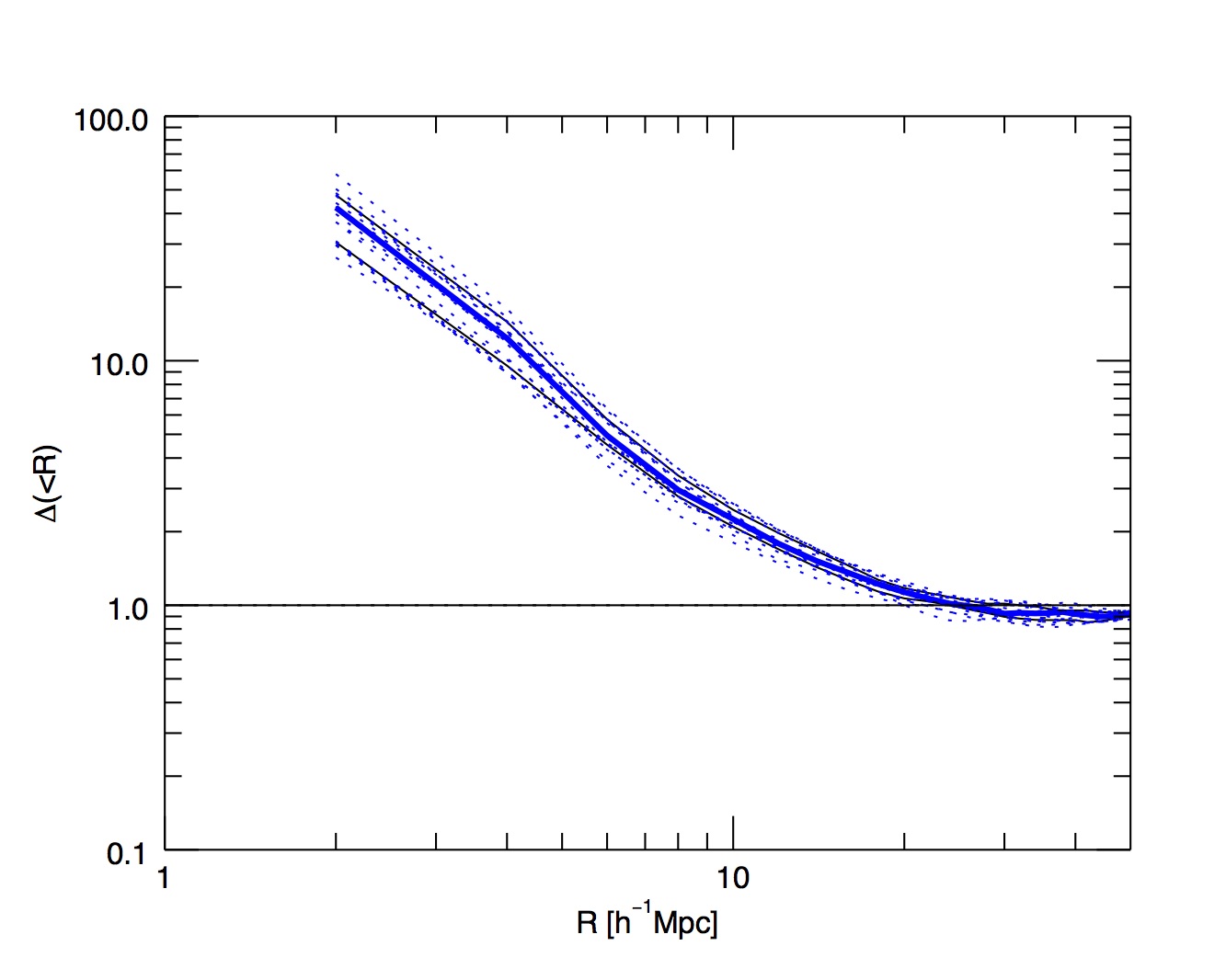}  
\hskip -0.3cm
\includegraphics[width=0.5\textwidth,angle=-00]{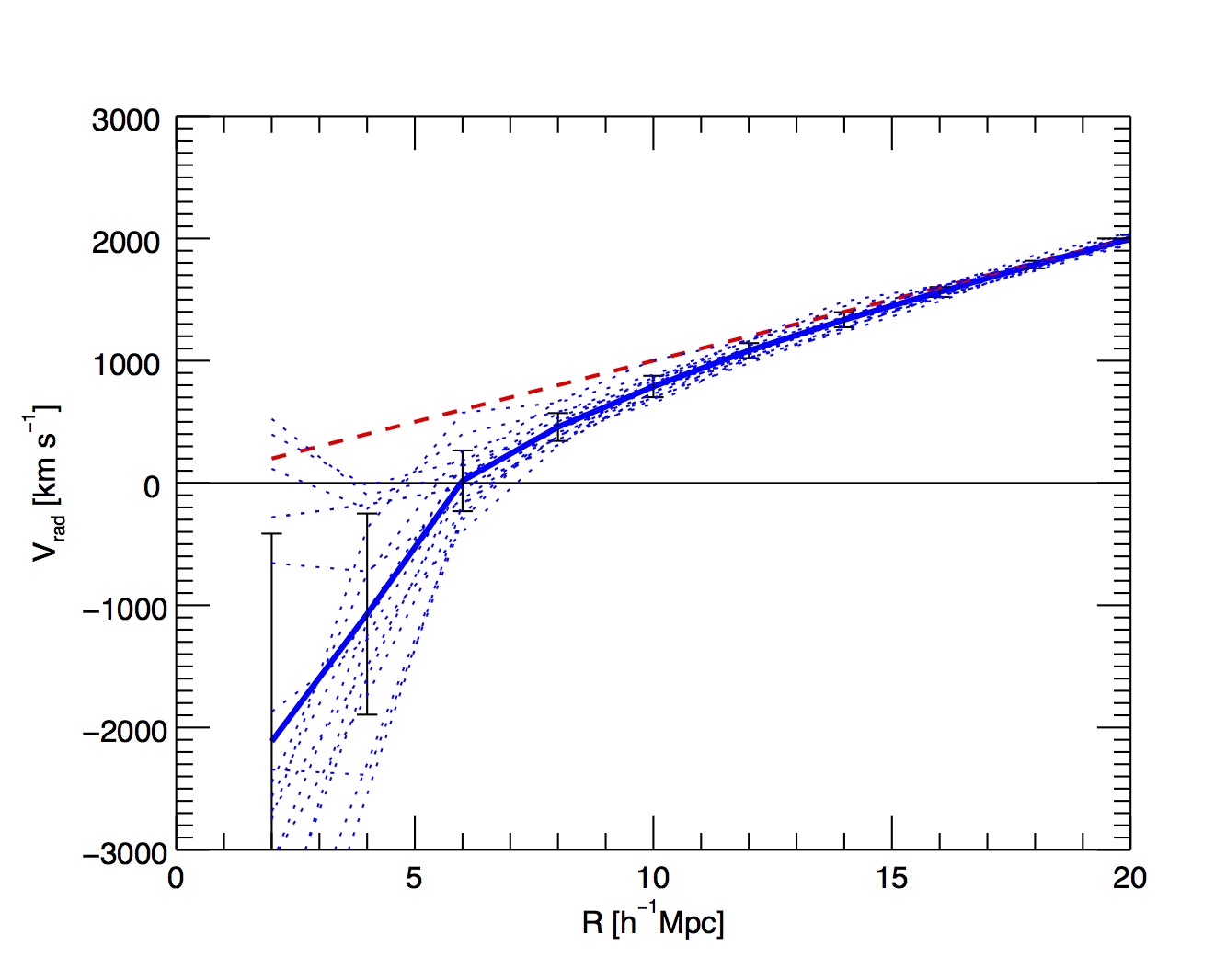}
\newline
\bf{Figure  7
}
\end{figure}


\clearpage

\newpage
{\bf Supplementary material}

{\bf Video:}\newline
The linked video   (\href{url}{http://vimeo.com/pomarede/quasilinear})  
starts with a representation of the linear WF estimated fractional over-density $\delta$ field.   The scene is within a box extending $\pm 80 \hmpc$  from the origin in the Supergalactic coordinate system.   The Milky Way lies at the apex of the  arrows of length $50 \hmpc$, indicating the positive supergalactic SGX (red), SGY (green), and SGZ (blue) directions. The linear $\delta$ field is represented by 6  isosurfaces :  $\delta=0.4$ in grey, and $\delta = 0.7,1.0,1.5, 2.0, 2.5 $ in nuances of red. Labels identify the major clusters and superclusters at local density maxima in red. The three-dimensional structure is explored by a sequence of rotations. At   0:28 the linear density field is replaced by the QL density, Gaussian smoothed at $R_s=4.0\hmpc$, at isosurfaces that correspond to $\Delta=1.2$ in grey, and $\Delta= 1.7, 2.0, 2.3, 2.7, 3.0 $ in nuances of red. The same sequence of rotations is repeated for the QL density.  
At 0:53,
there is a direct comparison of the density fields from the linear WF, left, and the QL, right.  The same large scale structure is uncovered by the two density fields, but the QL density map reveals a wealth of small scale structure that is absent from the linear, much smoother linear $\delta$ field.
At 1:26,
the right panel still represents the QL density field but
now the left panel presents a bias-corrected 2M++ density field, with the same color coding as the QL density field. Both fields are Gaussian smoothed with $R_s \sim 5.6 \hmpc$. The grid lines in the left panel represent the 2M++ zone of avoidance. The video progresses through a sequence of rotations. The agreement between the QL and 2M++ $\Delta$ fields is not perfect, but there is no manifestation of a systematic offset between the two.

{\bf Sketchfab:}\newline
The linked Sketchfab (\href{url}{https://sketchfab.com/models/f1c89ad5d4884dcf9c94a01810ac1c4b}) 
provides a three-dimensional interactive visualization of the cosmography uncovered by the QL reconstruction. 
The box, $\Delta$ isosurfaces and the   central signpost are the ones used in the video for the QL density field.   Annotations indicate major density enhancements; selecting annotations from a list will take the observer to predefined stations. The observer can freely explore the scene by mouse control (left-click and drag for rotations, right-click and drag for translation, wheel-roll for zoom in and out) or by the following actions on touchscreens: one finger drag for rotation, pinch for zoom in and out, two fingers drag for translations. Center of rotation can be modified by double-clicking on any point. Table 2 T  lists all the    objects marked in the visualization. The table provides the positions of all objects that are  identified by local maxima of the QL density field. The position of the MW is associated with the origin of the Supergalactic coordinates system.

\newpage
{\bf  Supplementary Table 1: Objects in Sketchfab visualization (\href{url}{https://skfb.ly/6toT7})  and their estimated positions in the QL density field}
\newline
\begin{tabular}{clccc}
\hline
              &                & SGX [$\hmpc$] & SGY [$\hmpc$]  &  SGZ [$\hmpc$] \\
\hline
1  & Milky Way              &        0.0              &          0.0           &   0.0                   \\
2 & Virgo  cluster       & -4.9                     &  12.7                 &       1.0                \\
3 &  Great Attractor    & -36.1                   &   17.6                &      -4.9                \\
4 & Perseus-Pisces supercluster  & 43.9 &  -16.6               &      -21.5               \\
5 & Cen-Pup-PP filament\cite{2017ApJ...845...55P}         &              &                         &                            \\     
 6 & Indus cluster  & -34.2    &           -29.3                 &      28.3               \\     
 7 & Arch\cite{2017ApJ...845...55P}  &  -6.8&    -14.6  &        55.7 \\
8  &  Coma  cluster & -11.7&       73.2 &     -4.9 \\
9 &  Norma-Pavo-Indus filament &  & & \\
10 &  Arrowhead\cite{2015ApJ...812...17P} mini-supercluster &  14.6&        22.5 &      6.8   \\
11 & Perseus-Pisces-Coma filament\cite{2017ApJ...845...55P} & &  & \\
12 &  Funnel\cite{2017ApJ...845...55P} &  69.3 &      -50.8 &      -46.9 \\
13 &  Hercules cluster & -41.0 &   64.5 &        66.4 \\
14 &  Great Wall  & & & \\
15  & A2162 cluster &  -7.8 &      55.7 &        79.1\\
\hline               
\end{tabular}


\begin{figure}
\label{fig:xxx}
\hskip -2.cm
\includegraphics[width=.5\textwidth,angle=-00]{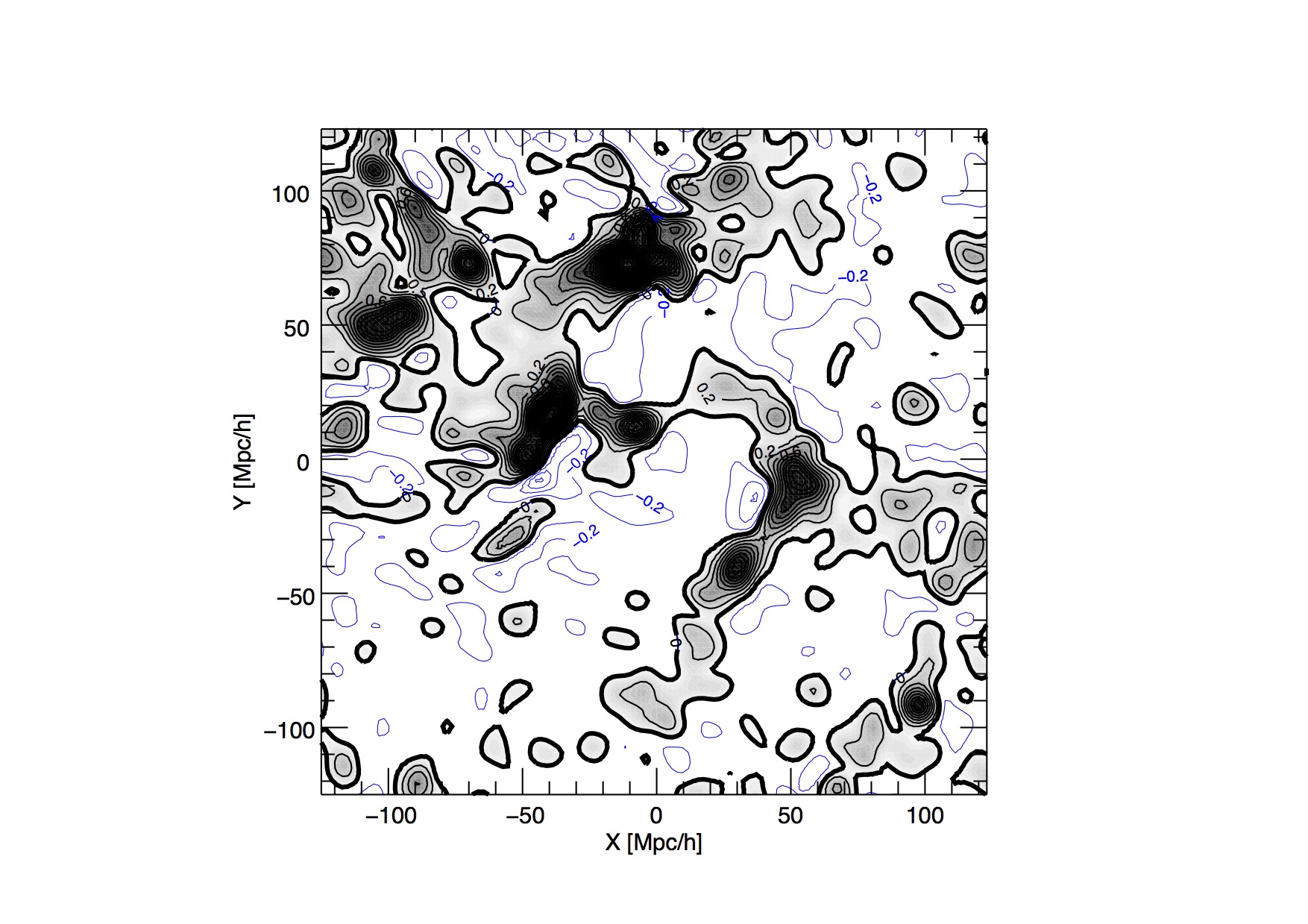}
\hskip -2.25cm
\includegraphics[width=.5\textwidth,angle=-00]{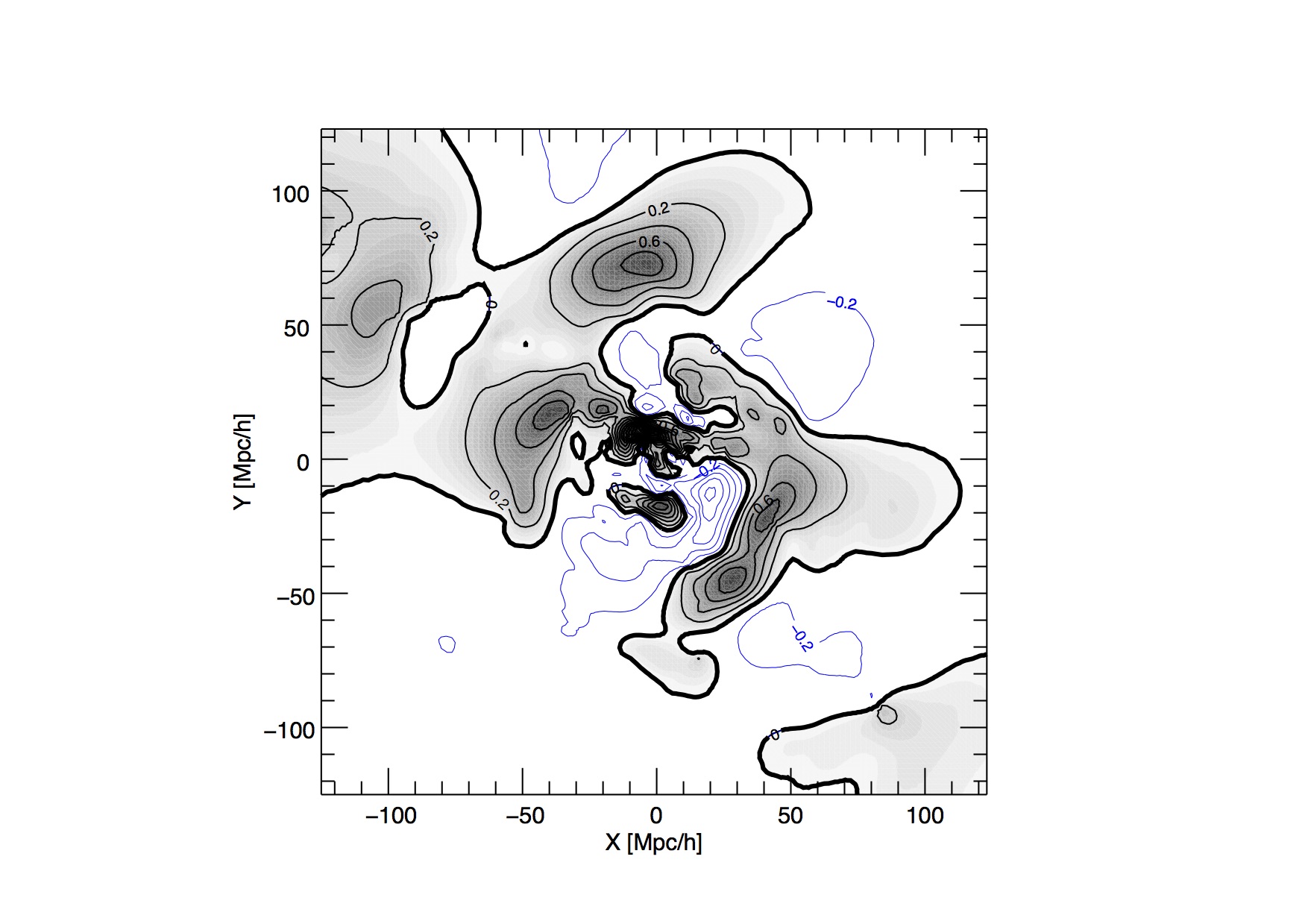}
\hskip -2.25cm
\includegraphics[width=.35\textwidth,angle=-00]{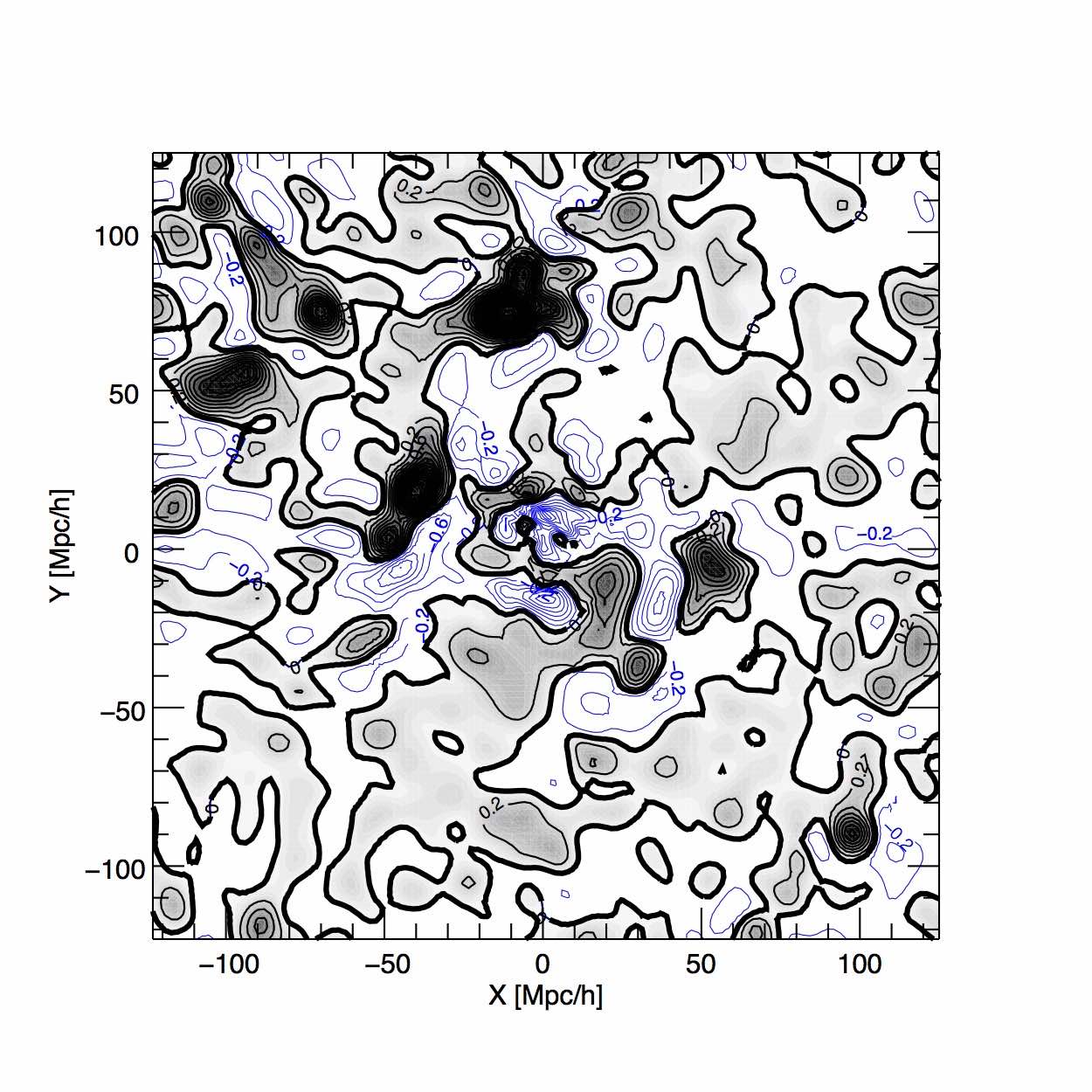}
\bf{Supplementary Figure 1:
Comparison of the scaled divergence of the velocity field ($-\nabla\cdot\bv/H_0$): the constructions with QL (left panel), linear WF (central panel), and the difference between the QL and WF fields (right panel). Contour spacing is 0.2, blue contours correspond to negative divergence and black contours to positive values.}
\end{figure}

\begin{figure}
\label{fig:SDvision}
\hskip -1.4cm
\includegraphics[width=.41\textwidth,angle=-00]{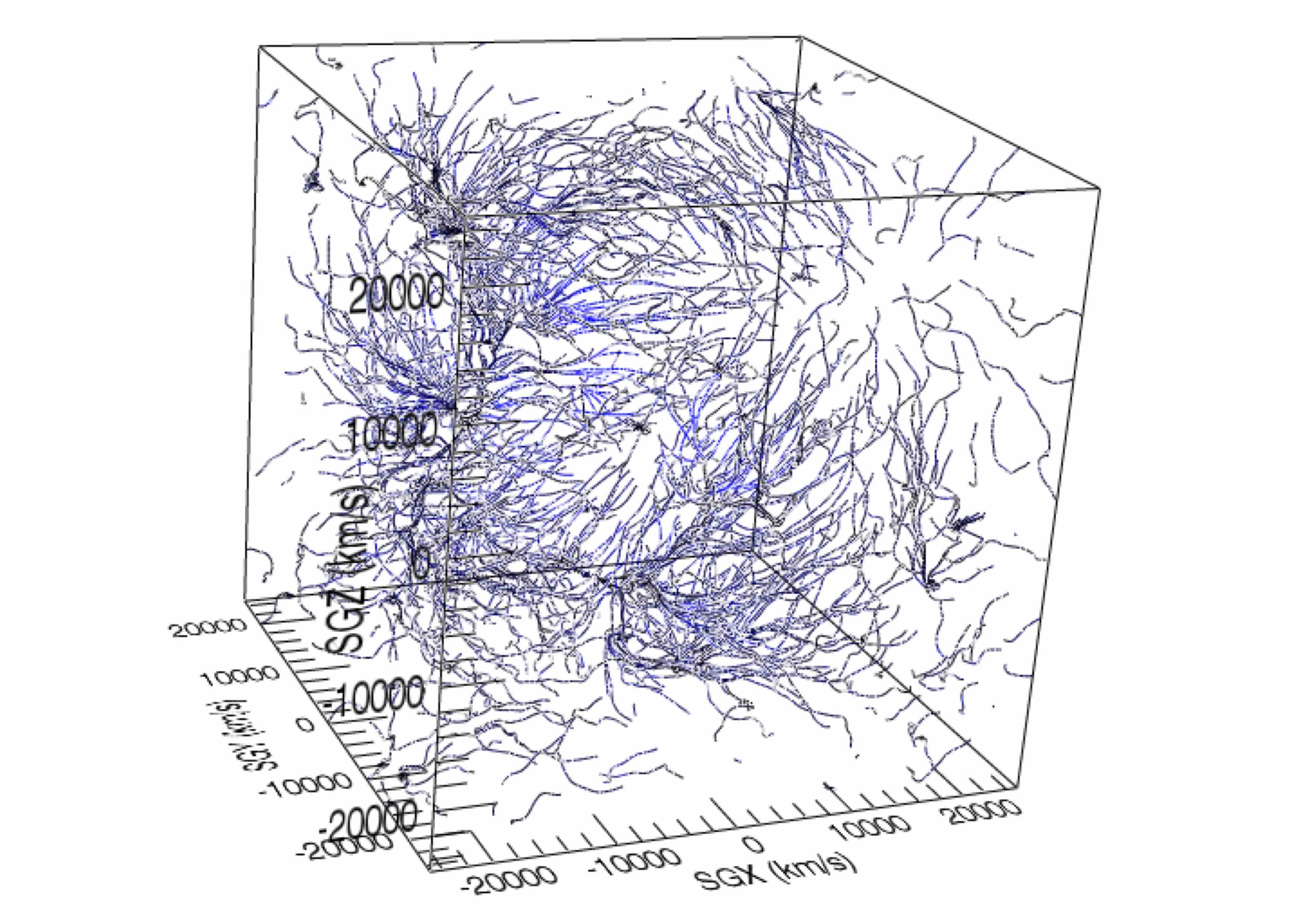}
\hskip -1.7cm
\includegraphics[width=.41\textwidth,angle=-00]{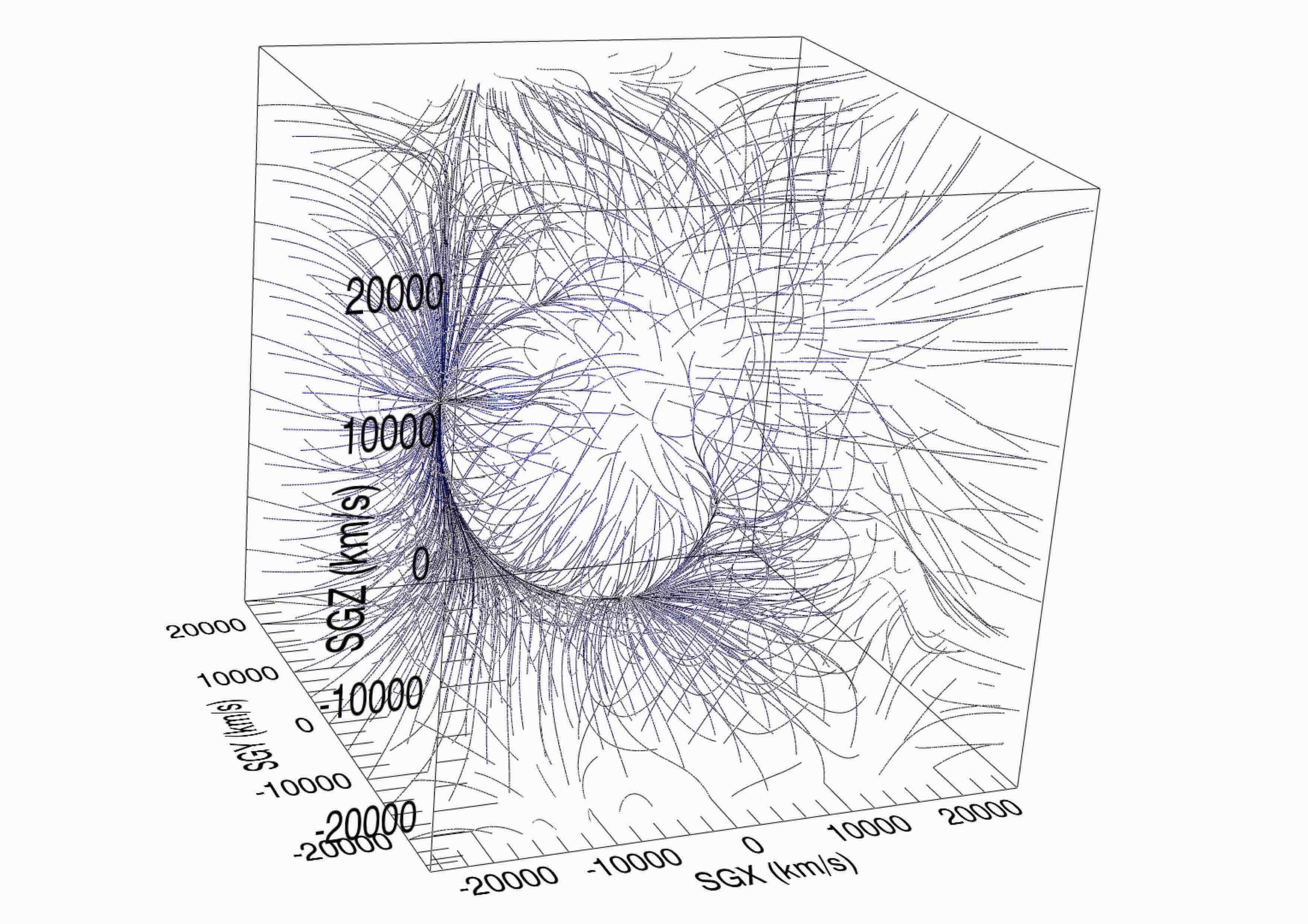}
\hskip -1.3cm
\includegraphics[width=.41\textwidth,angle=-00]{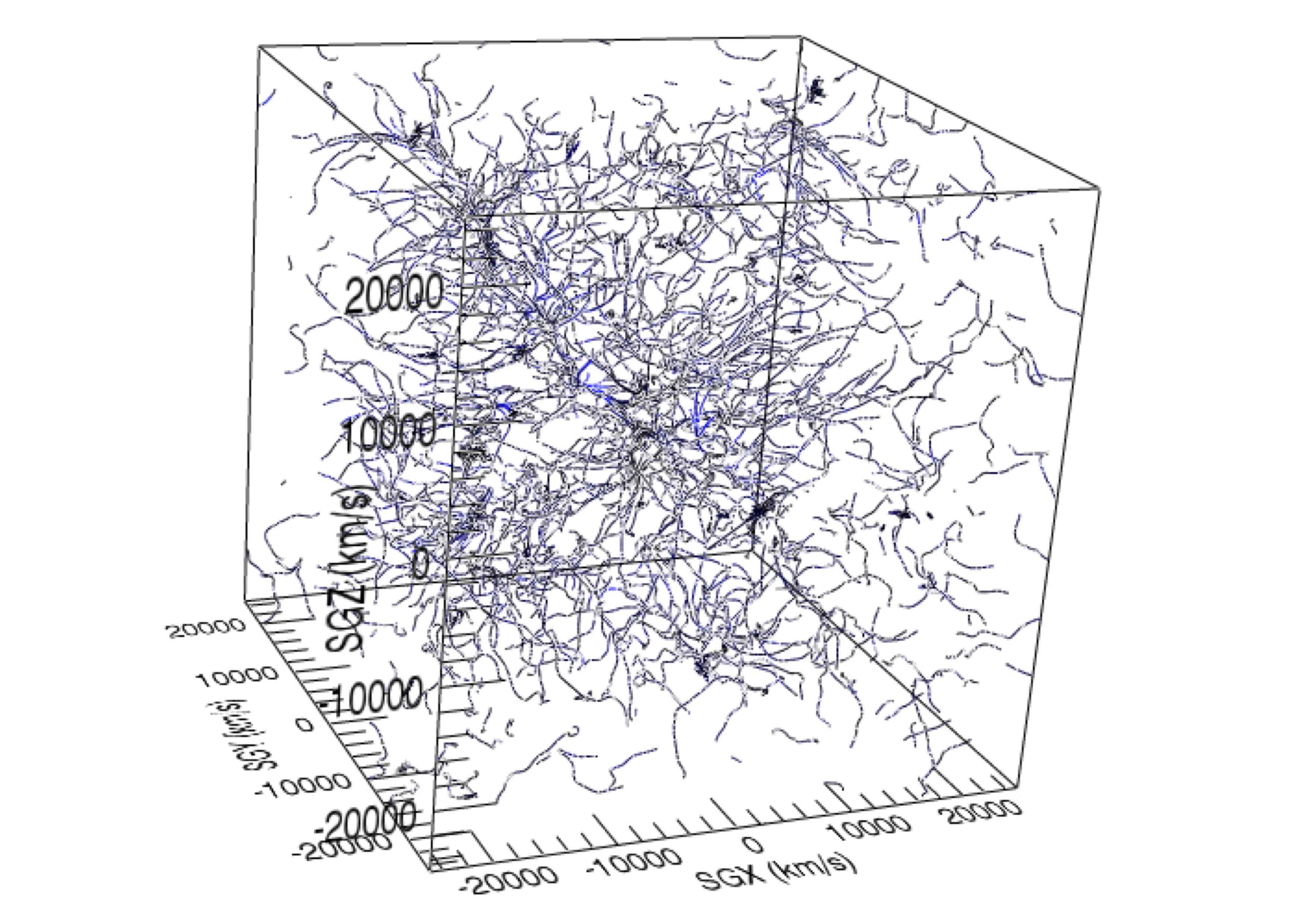}
\newline
\bf{Supplementary Figure 2: Comparison of the  velocity field  of  QL (left panel), linear WF (central panel), and the difference between the QL and WF fields (right panel).}
\end{figure}

\begin{figure}
\label{fig:S2N}
\hskip -.5cm
\includegraphics[width=.33\textwidth,angle=-00]{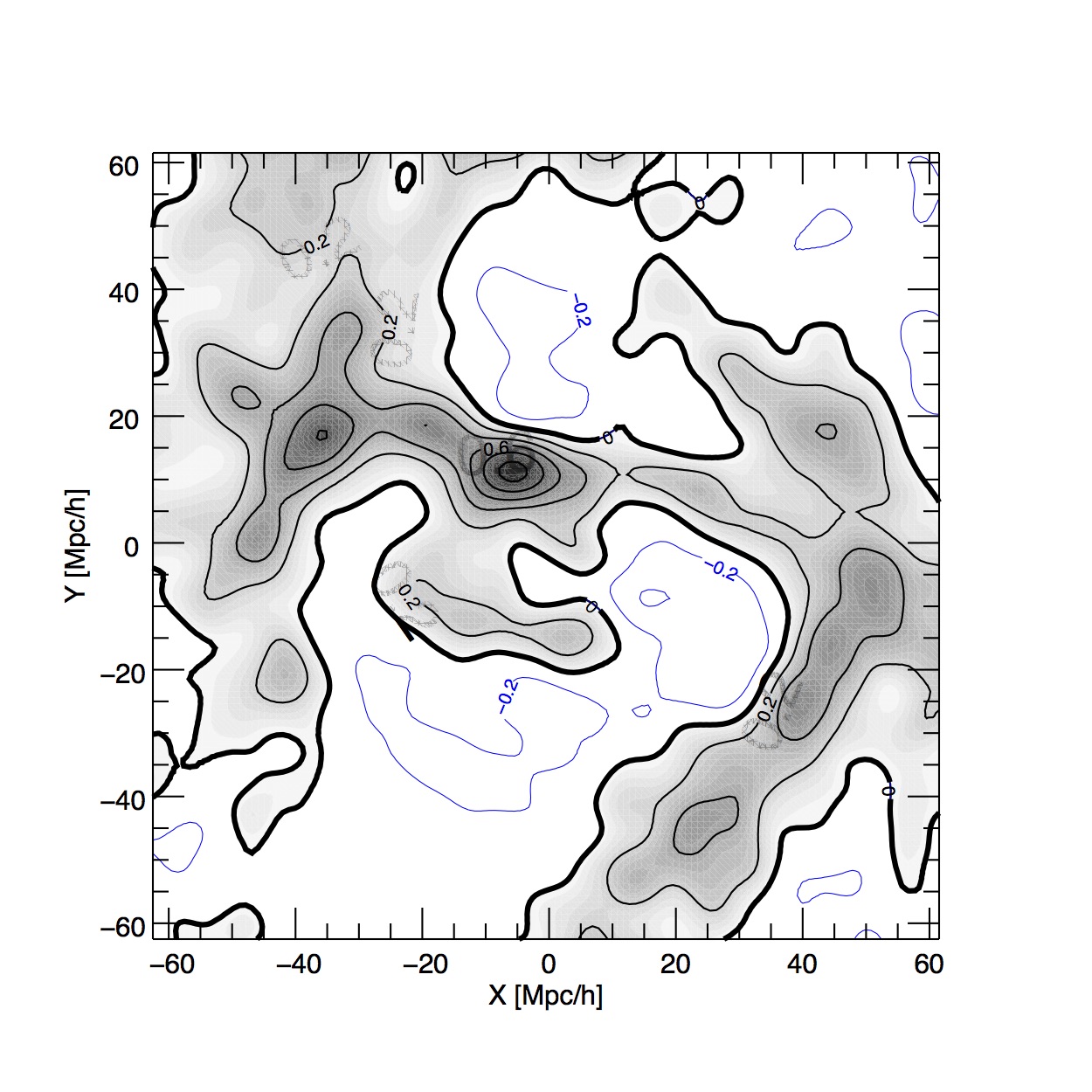}
\hskip -.5cm
\includegraphics[width=.33\textwidth,angle=-00]{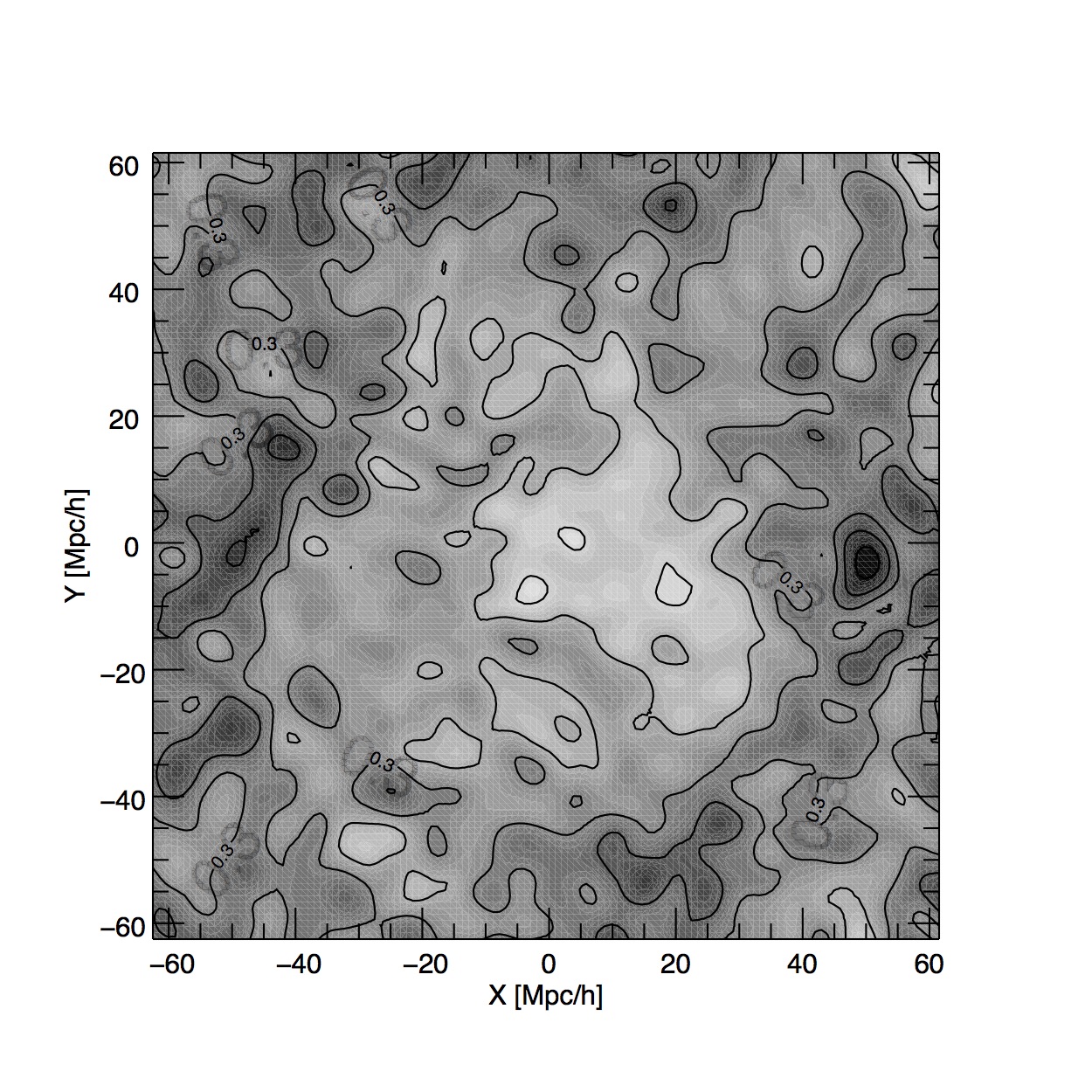}
\hskip -.5cm
\includegraphics[width=.33\textwidth,angle=-00]{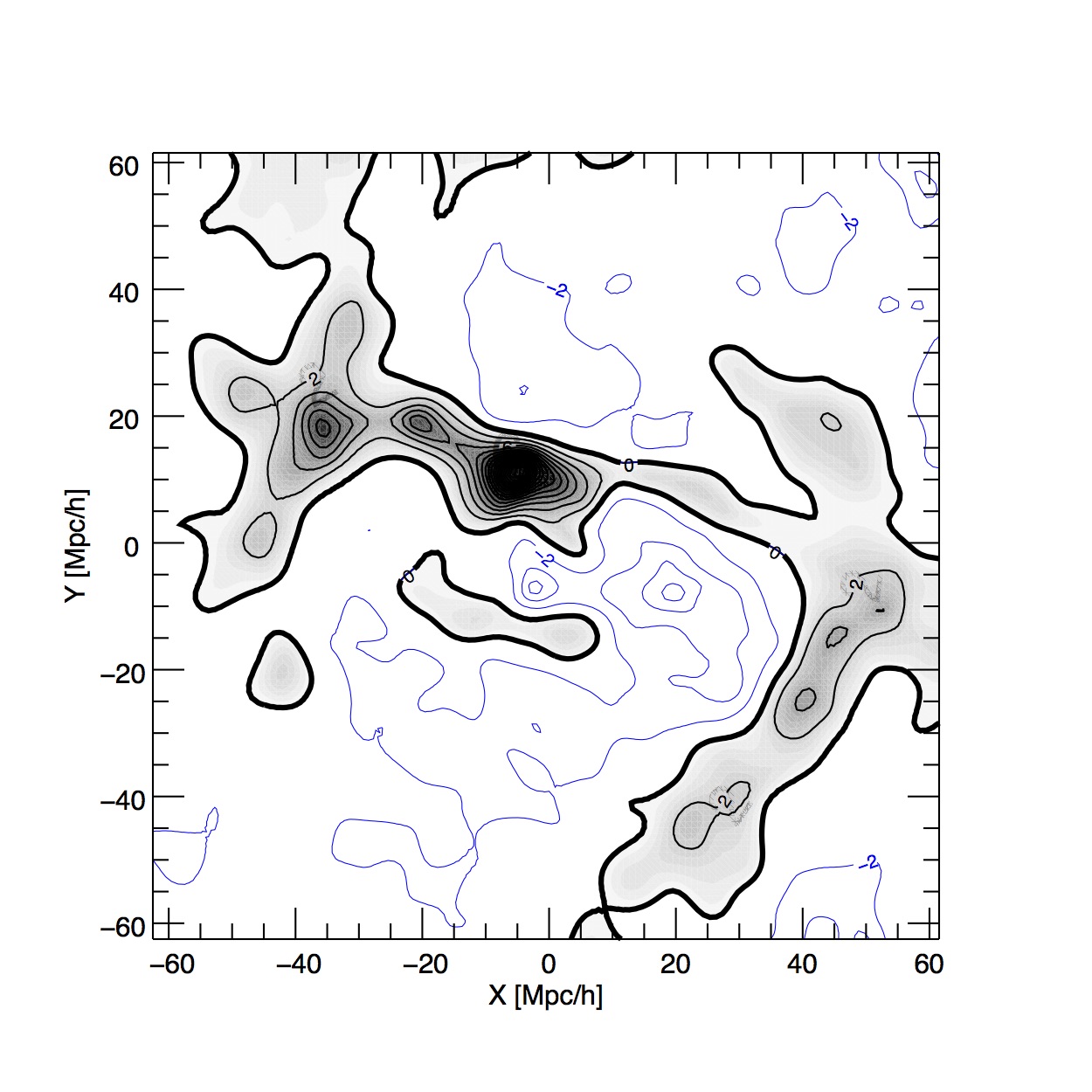}
\newline
\bf{Supplementary Figure 3: 
Contour maps of statistical uncertainties about the estimation of the QL density field. Left panel: the geometric mean values with contour spacing 0.2 of  $\log_{10}\Delta^{QL}$ and color convention following Figure 2. Central panel: the scatter around the mean field with contour spacing 0.1 in   $\sigma_\Delta$. Right panel: the signal-to-noise   contour map shows the   ratio of the mean to the scatter (i.e. signal to noise, S/N),  $(\Delta^{QL} - 1)/\sigma_\Delta$, with contour spacing 2.0.
}
\end{figure}

\newpage

\begin{figure}
\label{fig:aitoff-QL-2mrs-S2N-a}
\includegraphics[width=.4\textwidth,angle=-00]{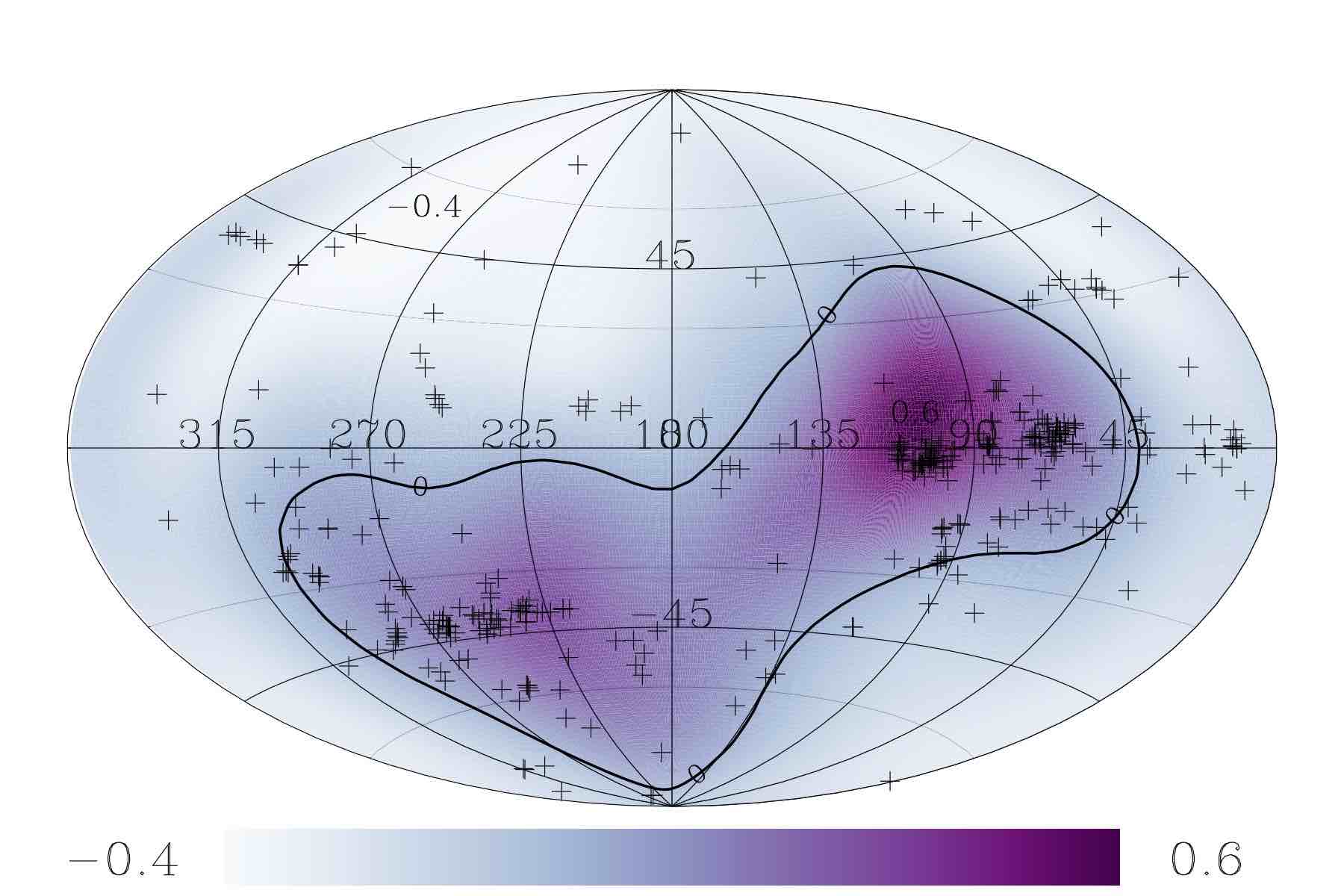} 
\includegraphics[width=.4\textwidth,angle=-00]{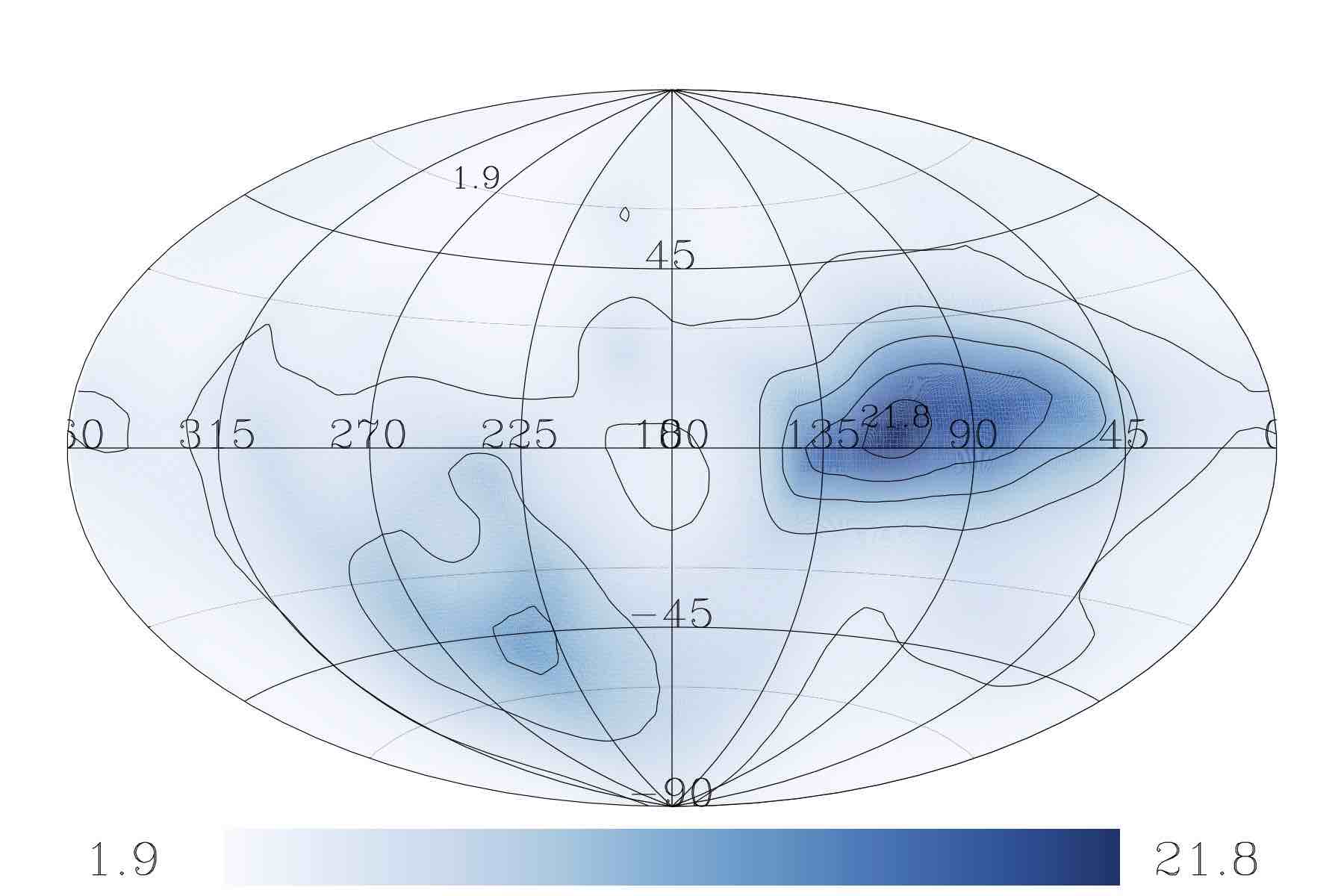} 
\newline 
\includegraphics[width=.4\textwidth,angle=-00]{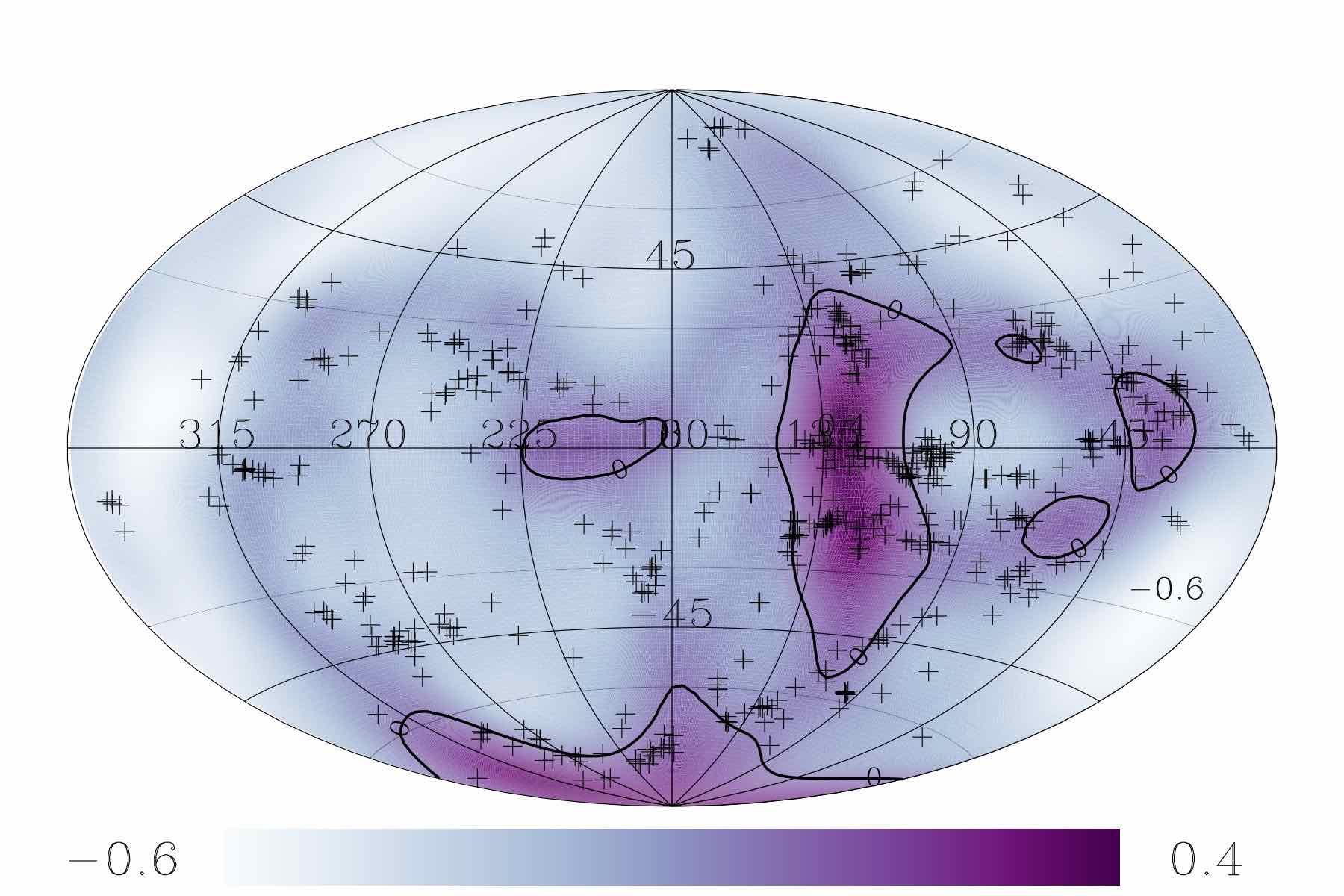} 
\includegraphics[width=.4\textwidth,angle=-00]{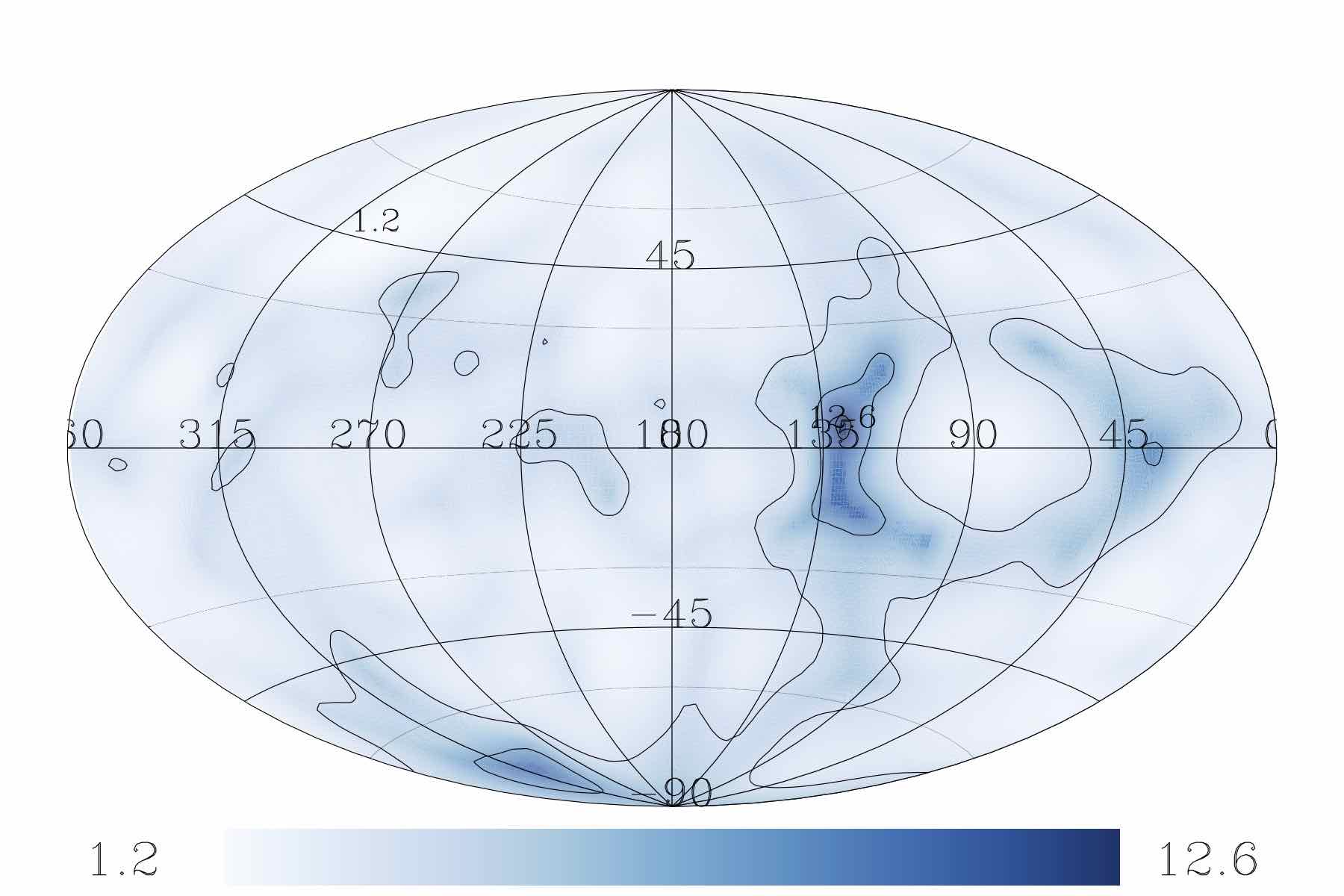} 
\newline 
\includegraphics[width=.4\textwidth,angle=-00]{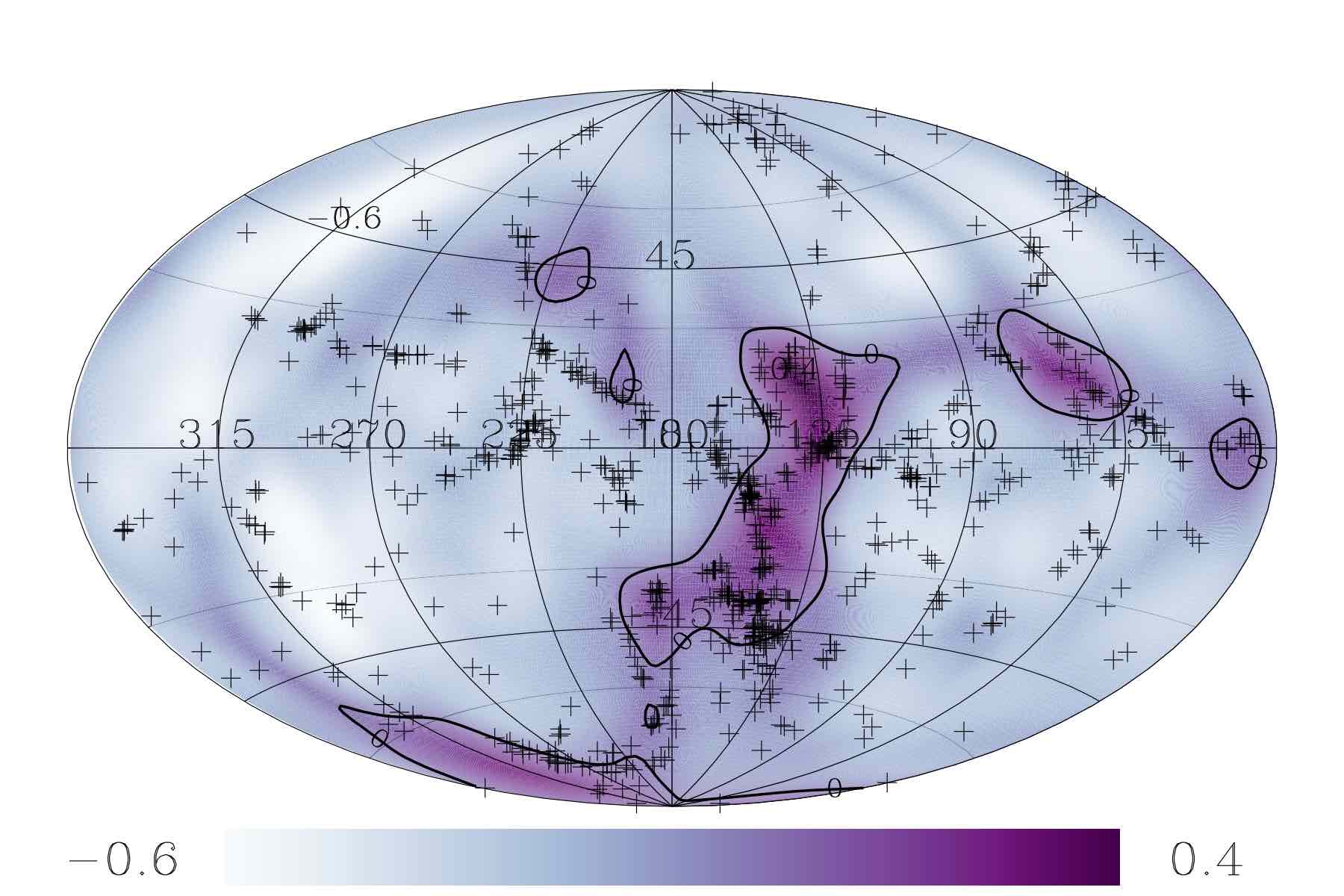} 
\includegraphics[width=.4\textwidth,angle=-00]{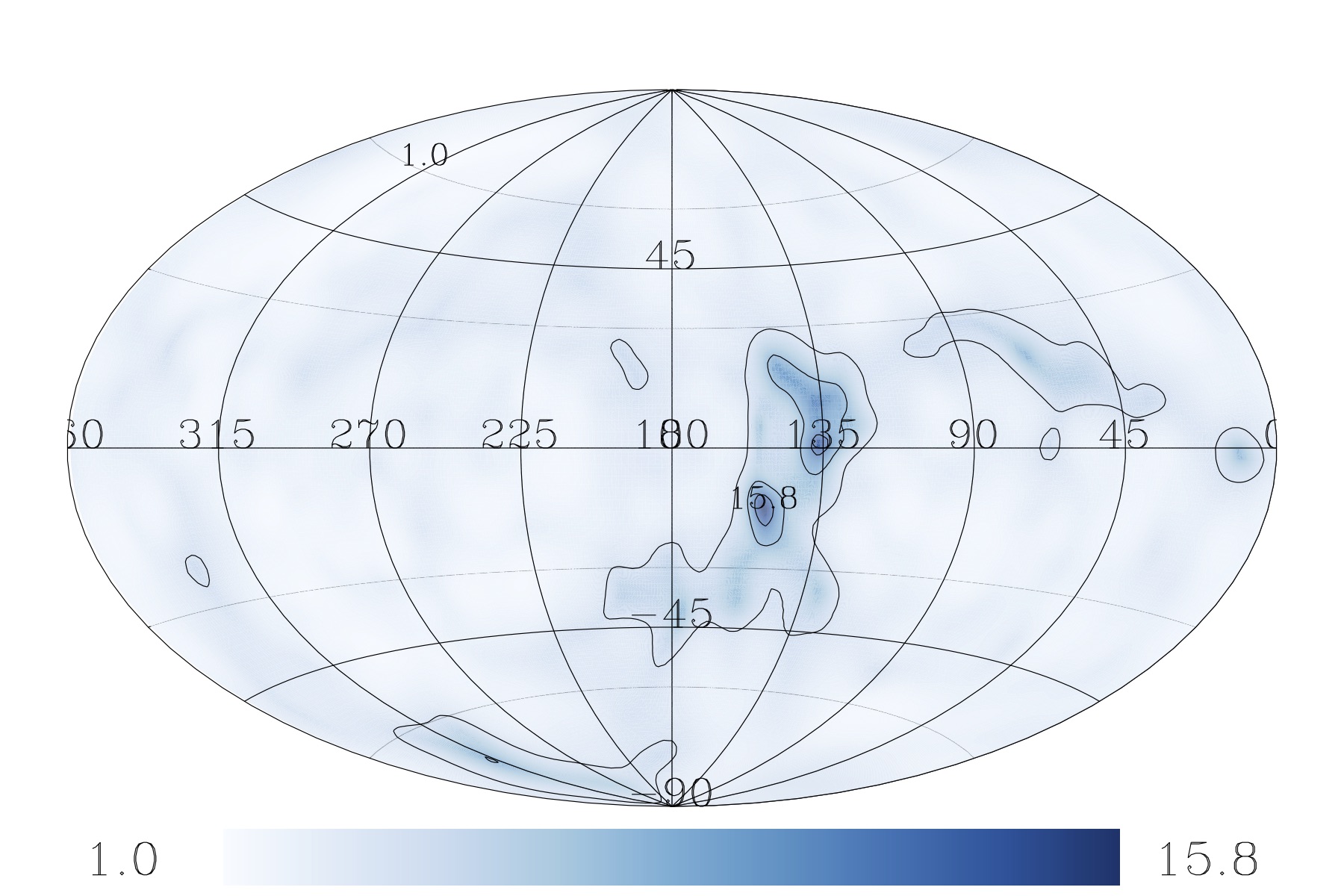} 
\newline 
\includegraphics[width=.4\textwidth,angle=-00]{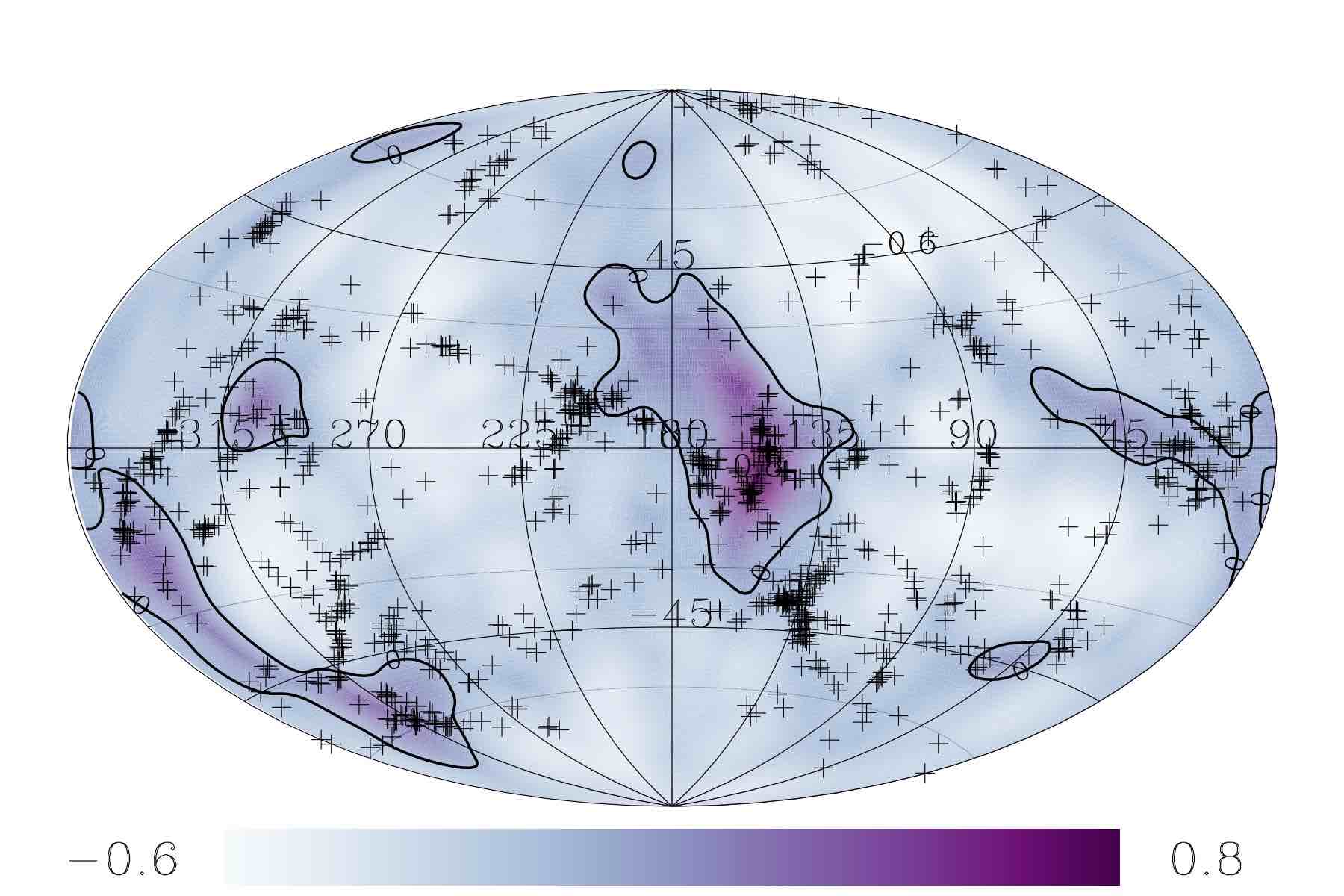} 
\includegraphics[width=.4\textwidth,angle=-00]{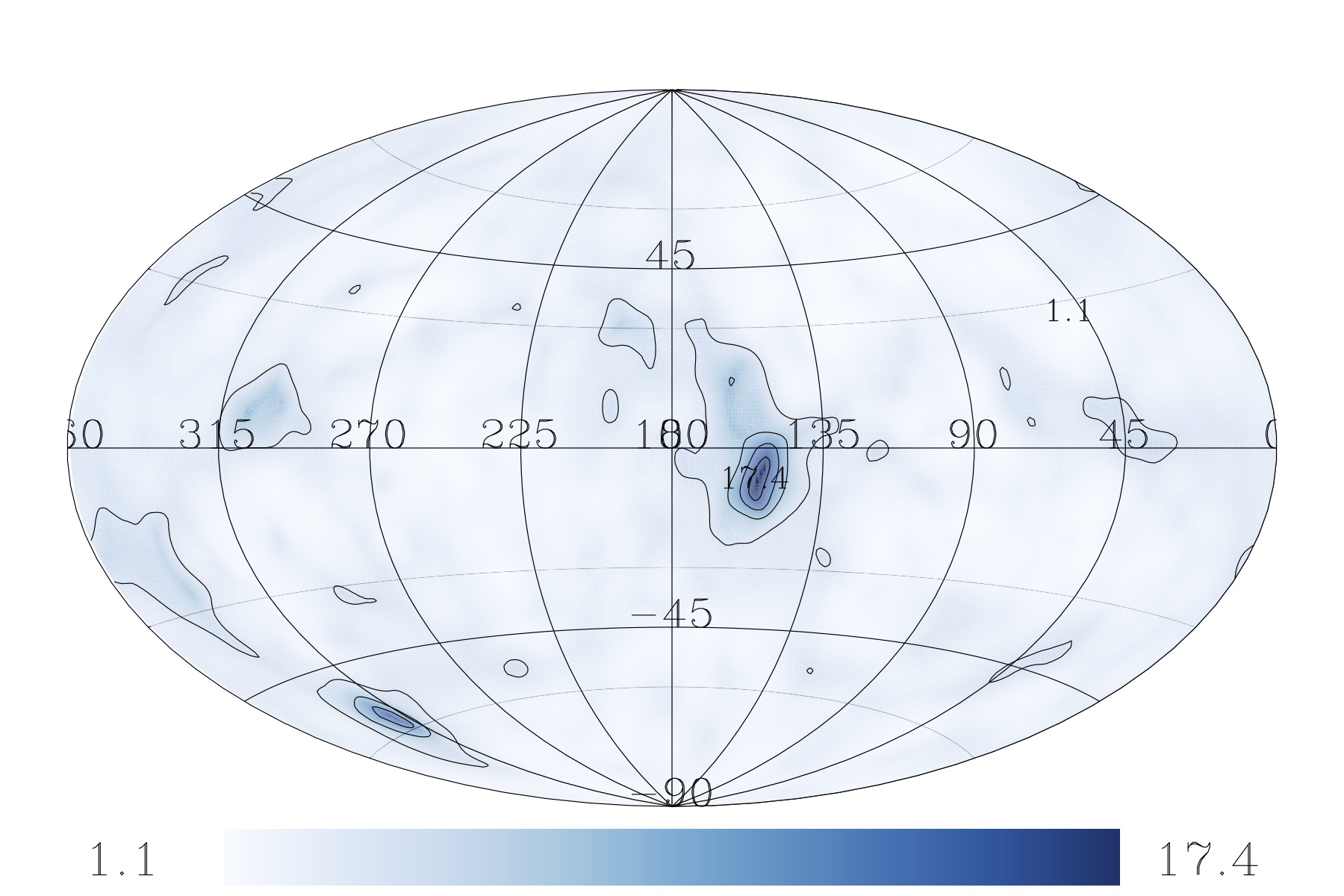} 
\newline 
\bf{Supplementary Figure 4a: 
Aitoff projections in supergalactic coordinates of the QL  ($\log_{10}\Delta$) density field  and  the 2M++ galaxies in  shells of R=10, ... 40$\hmpc$ (from top  to bottom) with   thickness of $\pm$5$\hmpc$ (left panels).  The right panels show   the (linear scale) signal-to-noise maps of the density field (contours spacing is 4.0., the values of the maximum and minimum S/N values are marked)
}
\end{figure}

\newpage
\begin{figure}
\label{fig:aitoff-QL-2mrs-S2N-b}
\includegraphics[width=.4\textwidth,angle=-00]{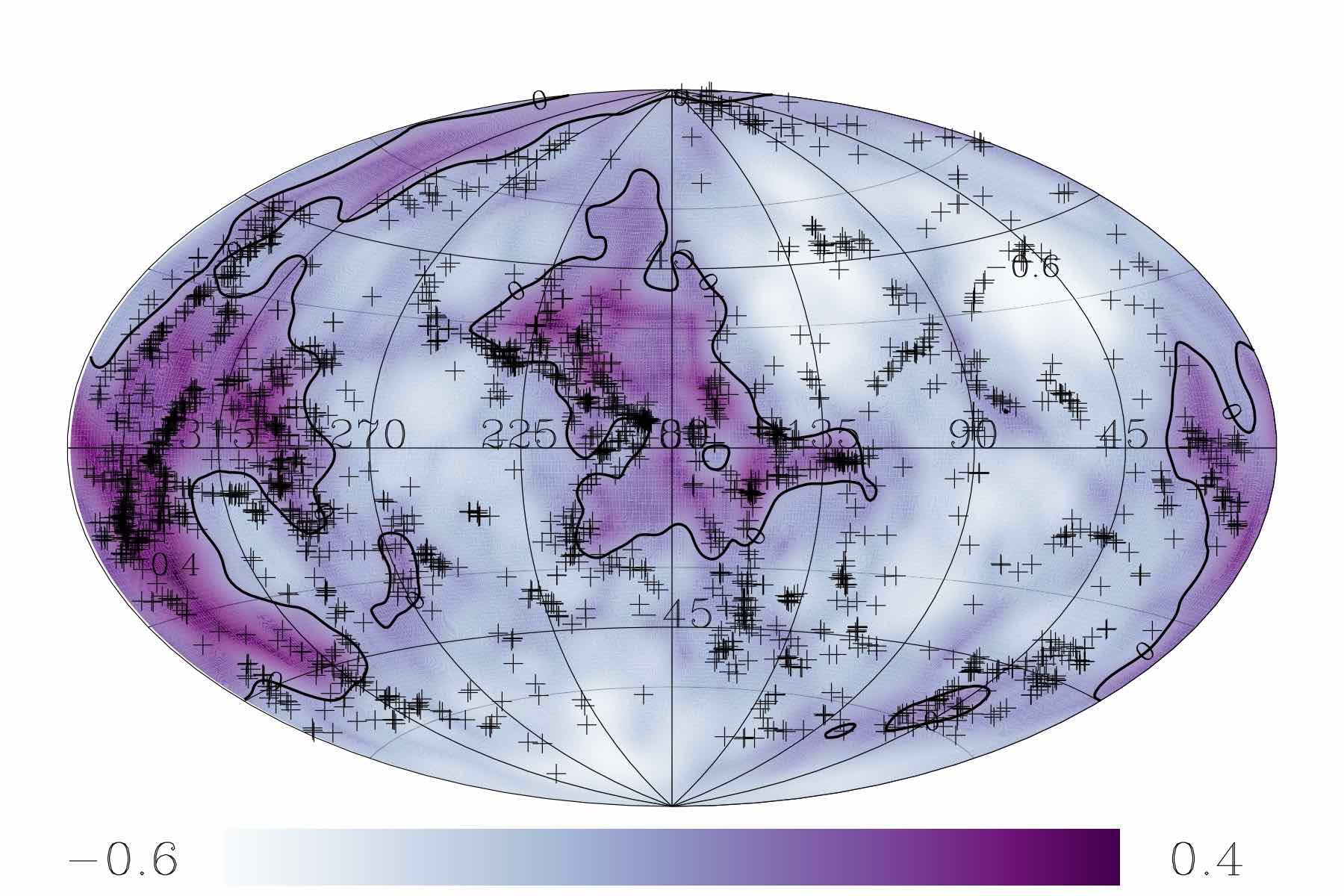} 
\includegraphics[width=.4\textwidth,angle=-00]{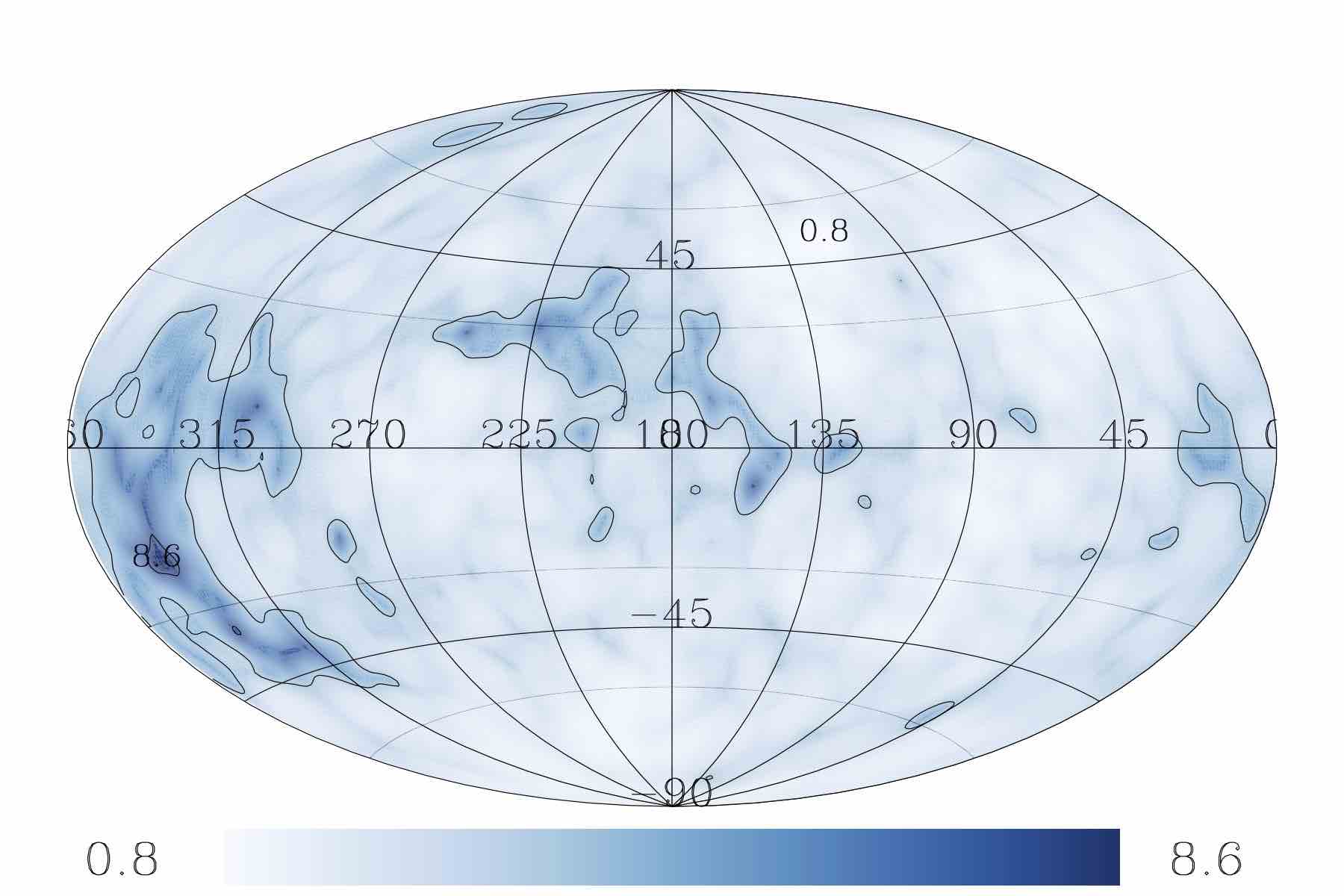} 
\newline 
\includegraphics[width=.4\textwidth,angle=-00]{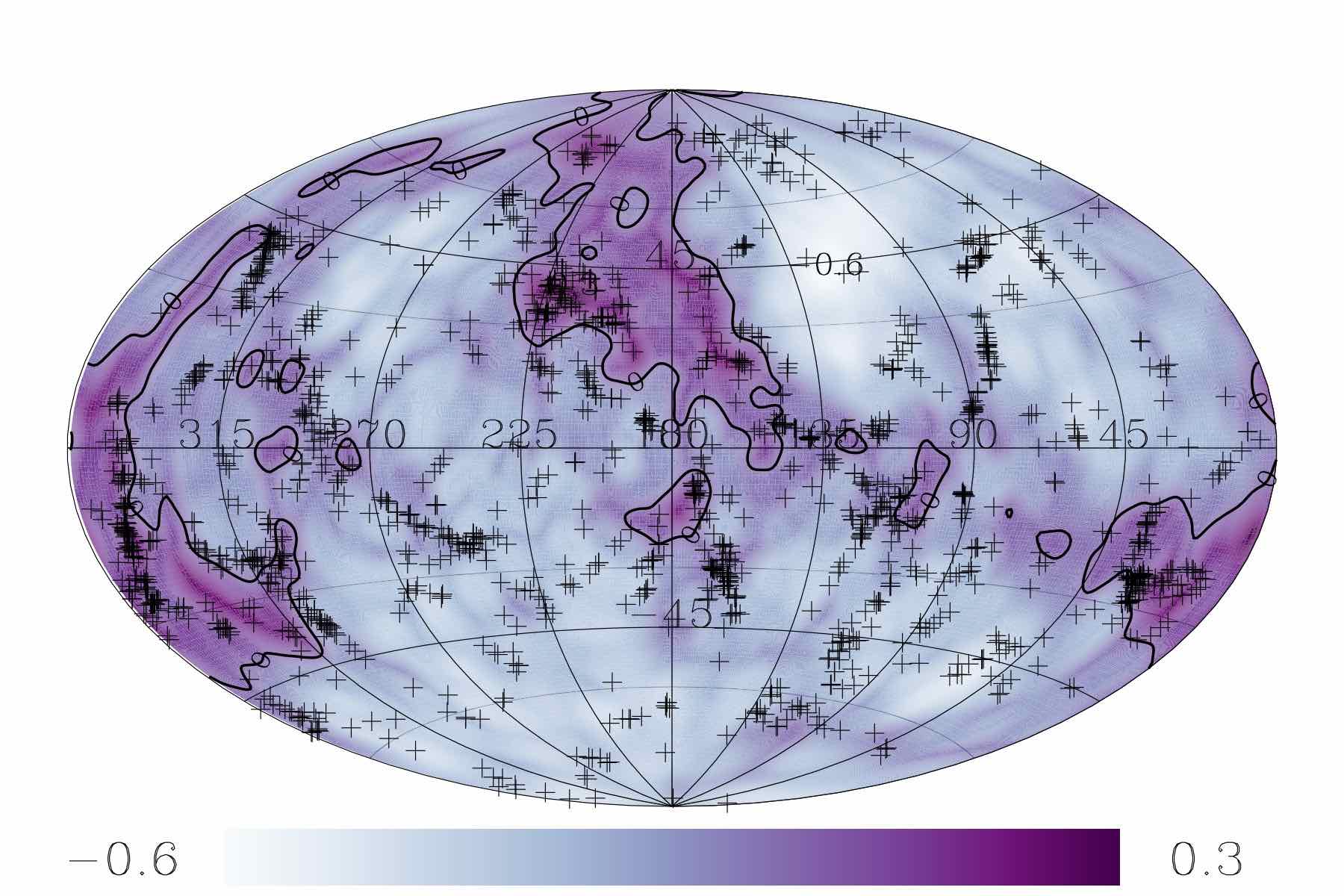} 
\includegraphics[width=.4\textwidth,angle=-00]{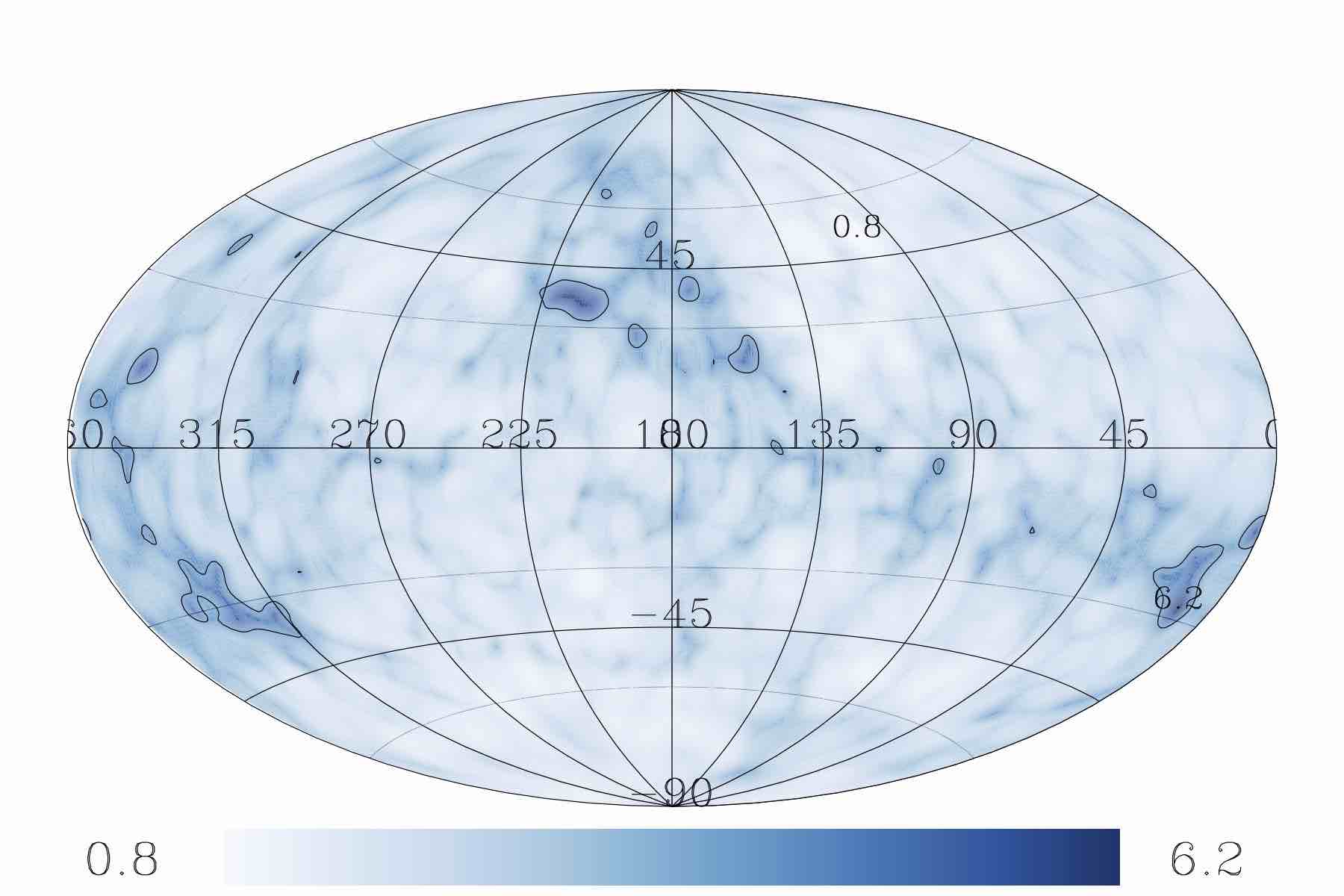} 
\newline 
\includegraphics[width=.4\textwidth,angle=-00]{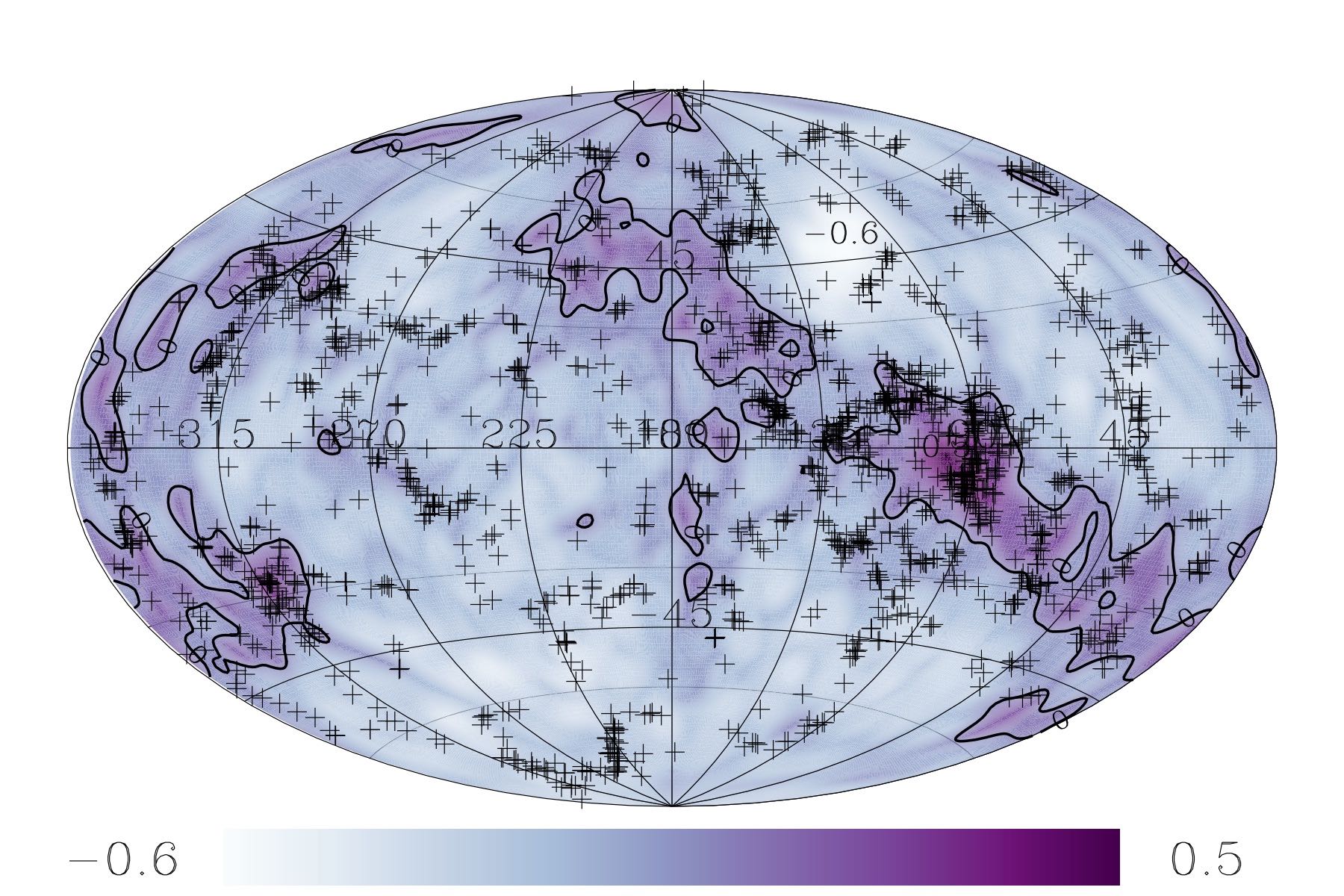} 
\includegraphics[width=.4\textwidth,angle=-00]{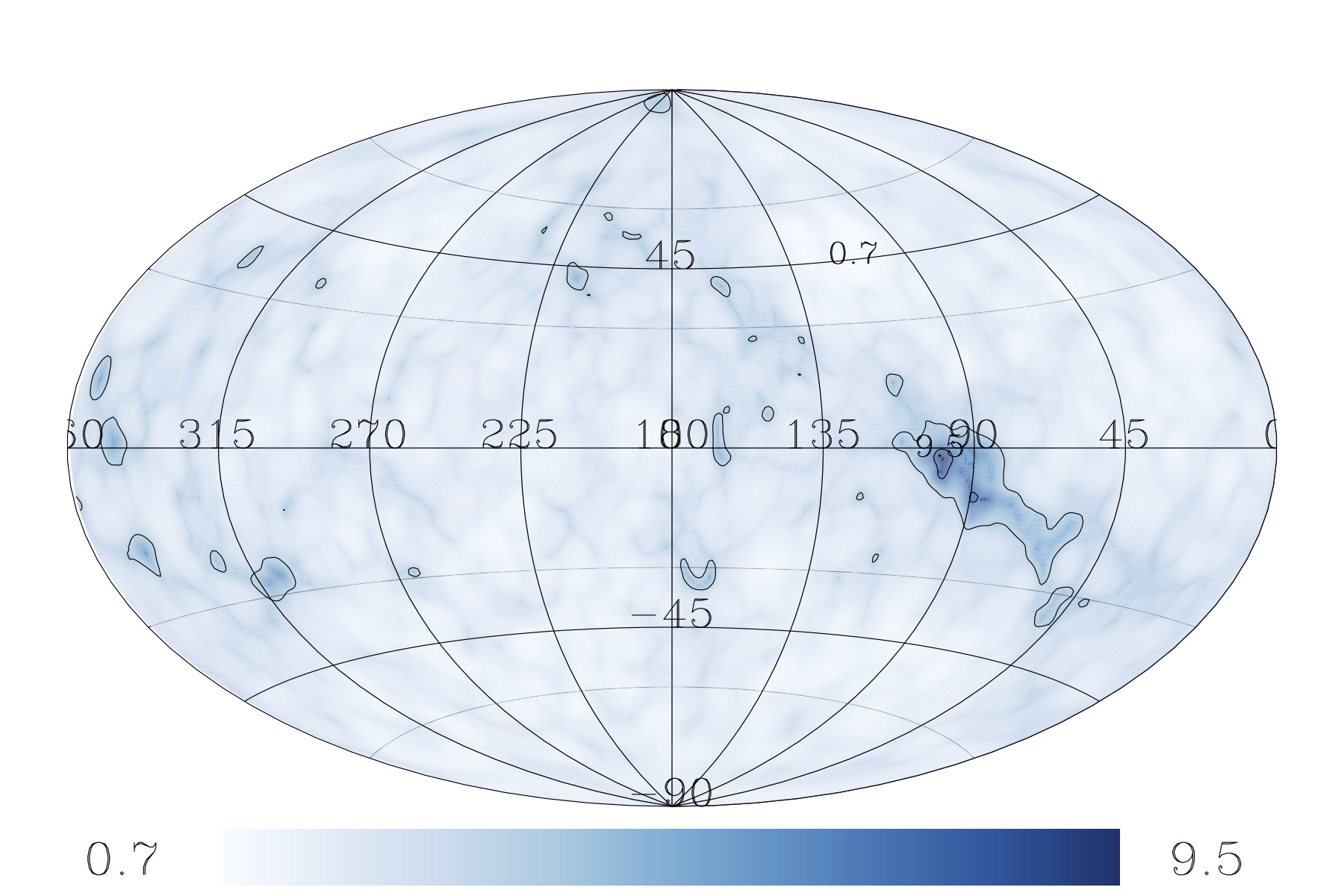} 
\newline 
\includegraphics[width=.4\textwidth,angle=-00]{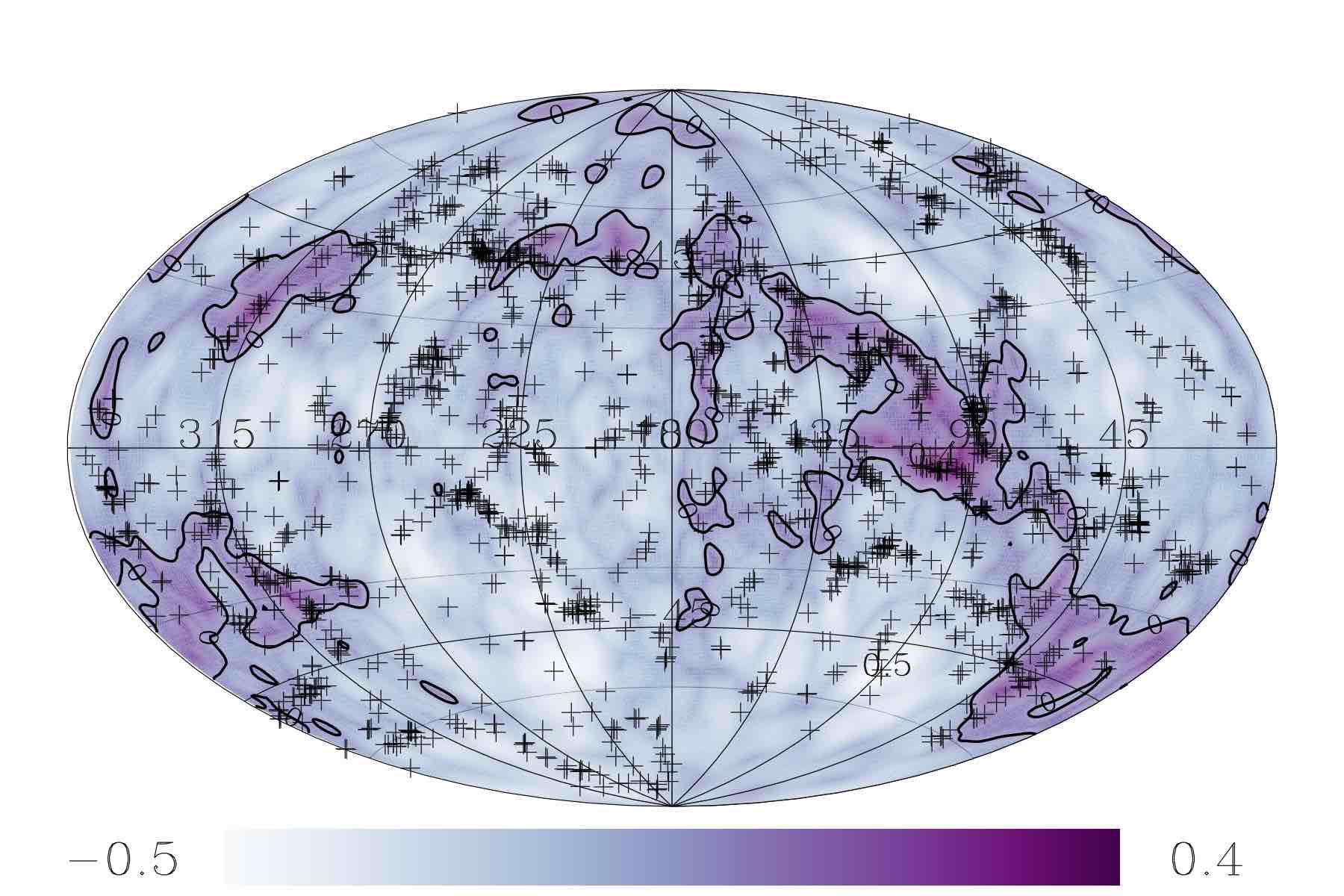} 
\includegraphics[width=.4\textwidth,angle=-00]{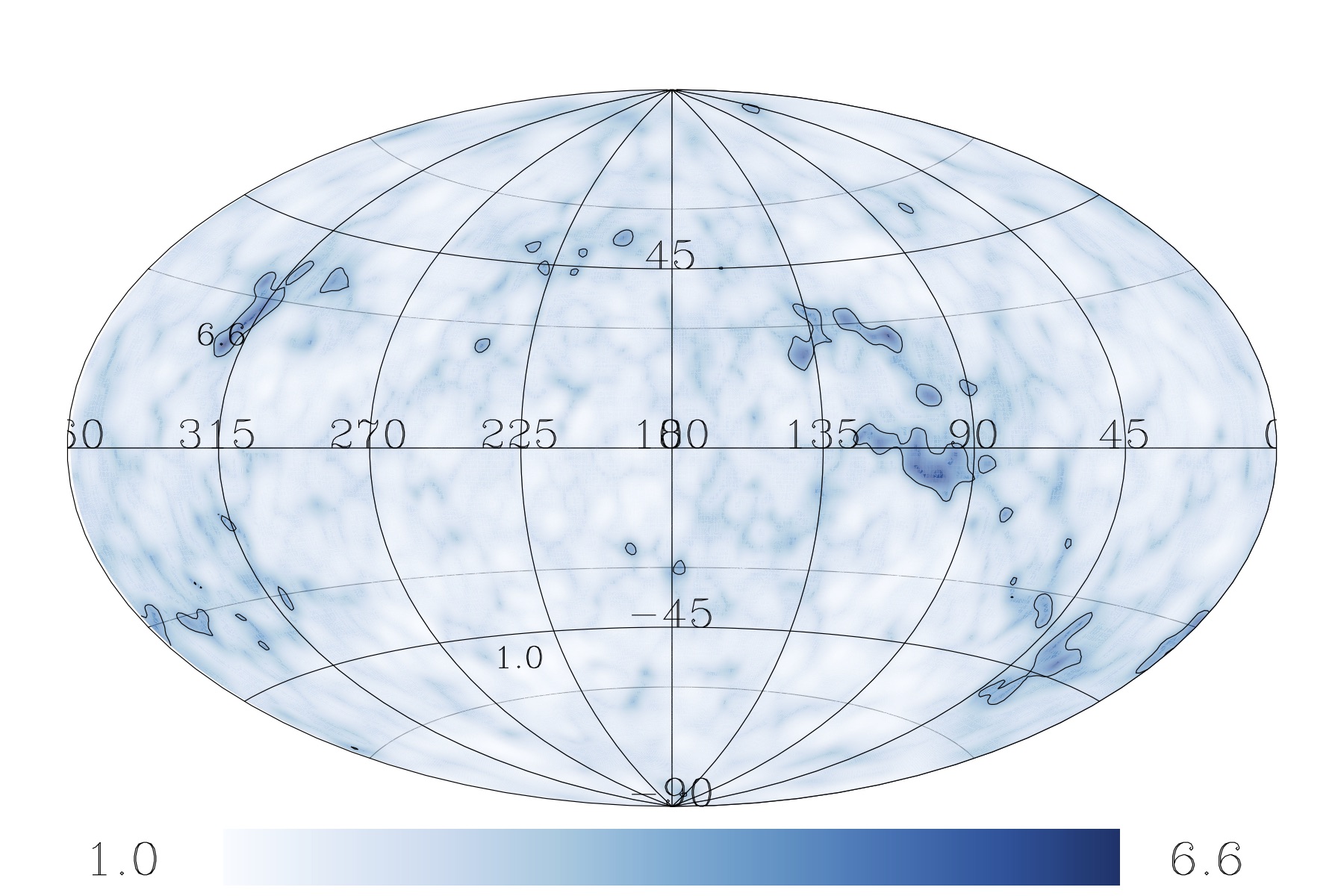} 

\bf{Supplementary Figure 4b: 
Same as Supplementary Figure 5a  but for R=50 (top), ... 80 (bottom)$\hmpc$.
}
\end{figure}

\end{document}